\documentclass[aps,prl,reprint,superscriptaddress]{revtex4-2}

\draft 
\usepackage{graphicx}
\usepackage{dcolumn}
\usepackage{bm}
\usepackage[separate-uncertainty = true, multi-part-units=single]{siunitx}
\usepackage[utf8]{inputenc}
\usepackage[T1]{fontenc}
\usepackage{mathptmx}
\usepackage{xcolor}
\usepackage{lineno}

\usepackage[normalem]{ulem}

\begin{document}


\title{Highly indistinguishable single photons from droplet-etched GaAs quantum dots integrated in single-mode waveguides and beamsplitters} 



\author{Florian Hornung}\thanks{Contributed equally to this work}
\email[]{f.hornung@ihfg.uni-stuttgart.de}

\author{Ulrich Pfister}
\thanks{Contributed equally to this work}

\author{Stephanie Bauer}
\thanks{Contributed equally to this work}
\affiliation{Institut für Halbleiteroptik und Funktionelle Grenzflächen, Center for Integrated Quantum Science and Technology (IQ$^{ST}$) and SCoPE, University of Stuttgart, Allmandring 3, 70569 Stuttgart, Germany}

\author{Dee Rocking Cyrlyson's}
\affiliation{Institute of Semiconductor and Solid State Physics, Johannes Kepler University Linz, 4040 Linz, Austria}

\author{Dongze Wang}

\author{Ponraj Vijayan}
\affiliation{Institut für Halbleiteroptik und Funktionelle Grenzflächen, Center for Integrated Quantum Science and Technology (IQ$^{ST}$) and SCoPE, University of Stuttgart, Allmandring 3, 70569 Stuttgart, Germany}

\author{Ailton J. Garcia Jr}

\author{Saimon Filipe Covre da Silva}
\affiliation{Institute of Semiconductor and Solid State Physics, Johannes Kepler University Linz, 4040 Linz, Austria}

\author{Michael Jetter}

\author{Simone L. Portalupi}
\affiliation{Institut für Halbleiteroptik und Funktionelle Grenzflächen, Center for Integrated Quantum Science and Technology (IQ$^{ST}$) and SCoPE, University of Stuttgart, Allmandring 3, 70569 Stuttgart, Germany}

\author{Armando Rastelli}
\affiliation{Institute of Semiconductor and Solid State Physics, Johannes Kepler University Linz, 4040 Linz, Austria}

\author{Peter Michler}
\affiliation{Institut für Halbleiteroptik und Funktionelle Grenzflächen, Center for Integrated Quantum Science and Technology (IQ$^{ST}$) and SCoPE, University of Stuttgart, Allmandring 3, 70569 Stuttgart, Germany}


\date{\today}

\begin{abstract}
\noindent The integration of on-demand quantum emitters into photonic integrated circuits (PICs) has drawn much of attention in recent years, as it promises a scalable implementation of quantum information schemes. A central property for several applications is the indistinguishability of the emitted photons. In this regard, GaAs quantum dots (QDs) obtained by droplet etching epitaxy show excellent performances with visibilities close to one for both individual and remote emitters. Therefore, the realization of these QDs into PICs is highly appealing. Here, we show the first implementation in this direction, realizing the key passive elements needed in PICs, i.e. single-mode waveguides (WGs) with integrated GaAs-QDs, which can be coherently controlled, as well as beamsplitters. We study both the statistical distribution of wavelength, linewidth and decay times of the excitonic line of multiple QDs, as well as the quantum optical properties of individual emitters under resonant excitation. Here, we achieve single-photon purities as high as $1-\text{g}^{(2)}(0)=\num{0.929(9)}$ as well as two-photon interference visibilities of up to V$_{\text{TPI}}$=\num{0.939(4)} for two consecutively emitted photons.  
\end{abstract}


\maketitle 

\begin{figure}[t!]
\includegraphics{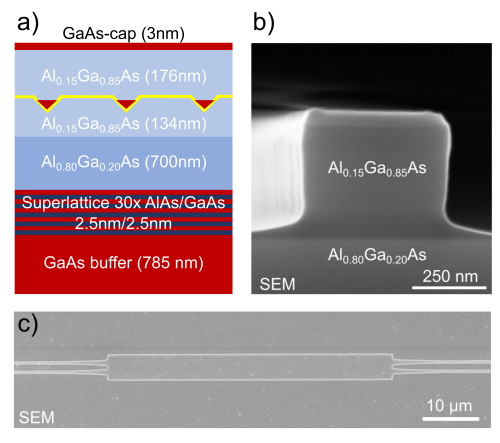}
\caption{a) Sketch of the layer structure of the WG-sample with GaAs QDs embedded in $\mbox{Al}_{0.15}\mbox{Ga}_{0.85}\mbox{As}$. b) SEM picture of a cleaved WG-facet. c) SEM top-view of an MMI-beamsplitter.}
\label{fig1}
\end{figure}

\begin{figure}[t!]
\includegraphics{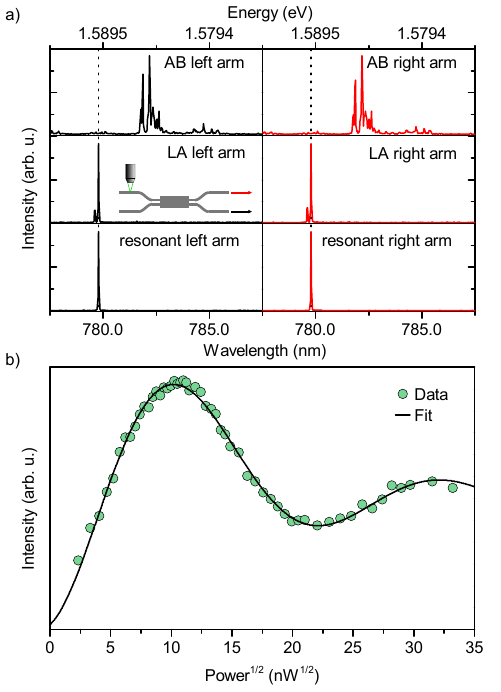}
\caption{a) Spectra of QD1 using different excitation wavelengths, including a \SI{532}{\nano \meter} above-band (AB) laser, pumping via longitudinal acoustic (LA) phonons and strictly resonant excitation. Photoluminescence was recorded at both output arms of the MMI, demonstrating the 50/50 splitting ratio of the device. b) Integrated intensity over excitation power for QD1 in resonant excitation with corresponding fit, showing clear Rabi oscillations.}
\label{fig2}
\end{figure}

\noindent Several quantum technologies such as optical quantum computing, quantum communication, quantum sensing and quantum simulation will strongly benefit from small-footprint photonic integrated circuits (PICs) due to the high scalability of the approach \cite{Wang.ea:2020,Moody.ea:2022}. A central part of these technologies are efficient on-demand sources of single and indistinguishable photons. In this regard, semiconductor quantum dots (QDs) are very promising candidates, showing very pure single-photon emission \cite{Schweickert.ea:2018}, high count rates \cite{Wang.ea:2019, Tomm.ea:2021, Nawrath.ea:2023} and emission linewidths close to the Fourier limit \cite{Kuhlmann.ea:2015, Pedersen.ea:2020, Strobel.ea:2023}. The compatibility of QDs with PICs has been demonstrated in various experiments, ranging from direct integration in fully monolithically grown samples \cite{Schwagmann.ea:2011, Laucht.ea:2012, Prtljaga.ea:2014, Kirsanske.ea:2017, Schnauber.ea:2018, Schwartz.ea:2018} to different hybrid integration techniques \cite{Zadeh.ea:2016, Davanco.ea:2017, Kim.ea:2017, Katsumi.ea:2018} and fiber-to-chip or chip-to-chip coupling \cite{Ellis.ea:2018, Singh.ea:2018, Bauer.ea:2021, Sund.ea:2023}. In order to increase the chip efficiency, different cavities compatible with waveguides (WGs) were introduced, which on the one hand increase the coupling efficiency into the WG and on the other hand shorten the decay time via the Purcell-effect, thus allowing higher repetition rates \cite{Arcari.ea:2014, Liu.ea:2018, Hepp.ea:2018, Dusanowski.ea:2020}. Another important point with respect to the up-scaling of quantum PICs is a tuning method to compensate the spectral mismatch between individual emitters and enable two-photon interference. Several tuning mechanisms have already been demonstrated for QDs in PICs, including local temperature tuning via laser heating \cite{Kim.ea:2018, Katsumi.ea:2020}, local \cite{Grim.ea:2019} and global strain \cite{Elshaari.ea:2018, Tao.ea:2020, Hepp.ea:2020} and electrical fields \cite{Petruzzella.ea:2018, Schnauber.ea:2021, Papon.ea:2023}. Nevertheless, interference between two WG-integrated remote QDs has been shown recently without an additional energy tuning knob \cite{Dusanowski.ea:2023}.\\
When having a closer look on different QD platforms, GaAs QDs obtained by droplet etching epitaxy (DEE) are very promising, as they show intrinsically short decay times due to their large lateral extent leading to excitons in the weak confinement regime \cite{Reindl.ea:2019}, near-unity two-photon interference visibilities for consecutive photons \cite{Reindl.ea:2019, Schoell.ea:2019} and a small wavelength distribution due to the strain-free growth \cite{daSilva.ea:2021}. Strain tuning \cite{Huber.ea:2018} and electrical tuning \cite{Huang.ea:2021} have also been demonstrated, as well as emission linewidths close to the Fourier-limit \cite{Zhai.ea:2020}. Furthermore, record-values in single-photon purity \cite{Schweickert.ea:2018} and in two-photon interference visibility of two remote sources \cite{Zhai.ea:2022} were achieved with DEE GaAs-QDs, but the integration of such QDs into PICs has not yet been shown.\\
Here, we demonstrate the on-chip guiding and splitting of single photons from DEE GaAs-QDs using a monolithic approach. The sample was grown via molecular beam epitaxy with a layer structure as shown in Figure~\ref{fig1}a. The necessary refractive index contrast for wave-guiding is achieved by different Al-contents in the two top layers (see supplementary for more information). A full-vectorial finite-difference eigenmode (FDE) solver was used to obtain the WG-dimensions for single-mode operation, giving a height of \SI[mode=text]{310}{\nano \meter} and a width of \SI[mode=text]{450}{\nano \meter} (see supplementary). The single-mode behavior of the WGs was confirmed by measuring the degree of polarization (DOP), resulting in an average of $\approx$~\SI{93}{\percent} (see supplementary). To simulate the ideal 2x2 multi-mode interference splitter (MMI) dimensions, the eigenmode expansion (EME) method was used, resulting in a length of \SI[mode=text]{58}{\micro \meter} with a width of \SI[mode=text]{4.5}{\micro \meter} (see supplementary). The WGs were processed using electron-beam lithography with a positive resist, followed by inductively coupled plasma reactive ion etching. To prevent oxidation, the etching was stopped about \SI[mode=text]{20}{\nano \meter} above the cladding layer. In a last step, the sample was cleaved perpendicular to the WGs and mounted in a helium-flow cryostat (operation temperature of \SI{4}{\kelvin}) with access from both the side and the top of the WGs for optical characterization. SEM-pictures of a cleaved WG facet and the top view of an MMI can be seen in Figure~\ref{fig1}b and c, respectively.\\
The sample consists of several MMIs, where the length was swept from \SIrange{57}{61}{\micro \meter} in \SI{1}{\micro \meter} steps, since the simulations predict a 50/50 splitting ratio with high transmission in this range. Additionally, straight reference WGs without MMI were also processed, which were used to measure an attenuation of \SI{8.15(108)}{\decibel \per \milli \meter} (in a wavelength range from \SIrange{740}{770}{\nano \meter}, see supplementary for details). The splitting ratio of the devices was verified by comparing the micro-photoluminescence emission at the two outputs of the MMI. Representative spectra of a QD located in an MMI with \SI{60}{\micro \meter} length are shown in Figure~\ref{fig2}a for above-bandgap excitation, as well as phonon-assisted and strict resonant excitation of the neutral exciton (X), demonstrating the 50/50 splitting into the two output arms. The transmission through the MMIs was determined to \SI{82.2(12)}{\percent} (see supplementary for more information). Note that the resonant spectra are recorded without spectral filtering, making use of the spatial separation of the excitation and the collection paths. Remaining laser stray light is additionally suppressed by coupling the light into a polarization-maintaining fiber and using a pulse-shaper to spectrally narrow the laser pulses (resulting in a pulse length of about \SIrange{20}{30}{\pico \second}) \cite{Schwartz.ea:2016}. Interestingly, the X transition can't be seen in above-band excitation for this specific QD, which we ascribe to additional charges due to the nearby WG-sidewalls and a weak background doping (see supplementary). The good laser suppression allows to observe clear Rabi-oscillations, as illustrated in Figure~\ref{fig2}b. The fit to the data gives an excitation fidelity of \SI{58.4(25)}{\percent}.\\       
\begin{figure*}[t!]
\includegraphics{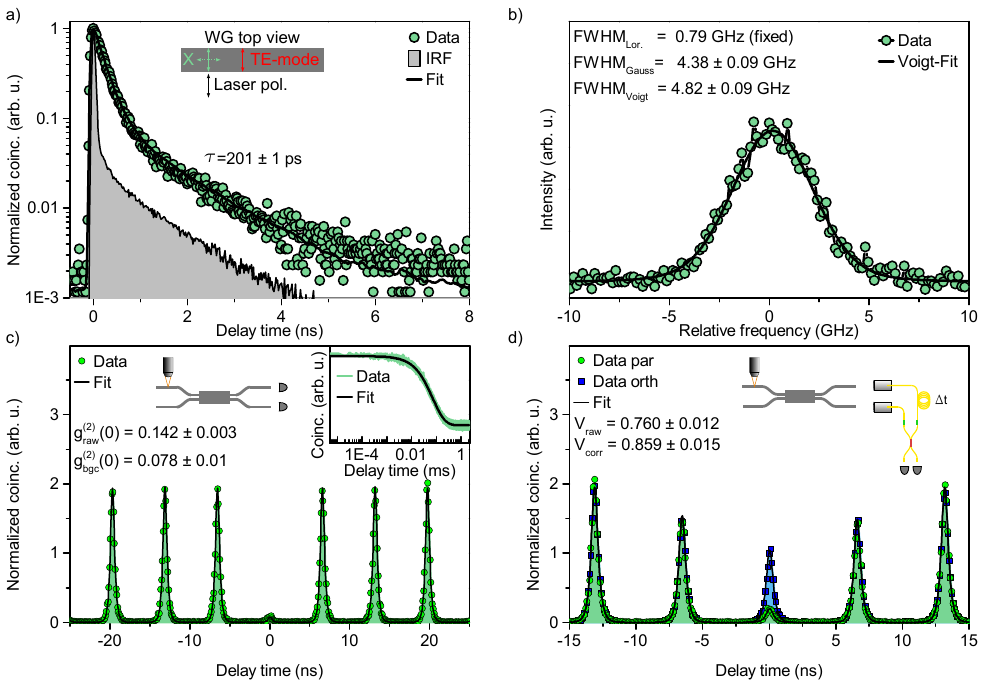}
\caption{a) TCSPC measurement for the X of QD1 under resonant excitation and corresponding fit including a deconvolution. The lack of a beating indicates that only one fine-structure component can couple to the WG-mode, as shown in the sketch. b) FPI-measurement of the X under resonant excitation and corresponding Voigt-fit. c) Second-order correlation measurement of the X of QD1 under pulsed resonant excitation using the on-chip beamsplitter. The fit to the data includes the deconvolution with the IRF. Inset: Correlation measurement obtained using a coarse bin size corresponding to one excitation cycle (\SI{6.57}{\nano \second}), plotted for delay times up to \SI{2}{\milli \second} and corresponding fit to extract the bunching parameters. d) Hong-Ou-Mandel measurement of the X of QD1, using the on-chip MMI to divide the photon stream and a fiber-beamsplitter to recombine it. The photons of one arm are delayed by the repetition rate of \SI{6.57}{\nano \second} and a polarization control (not shown in the sketch) allows to measure the co- and cross-polarized case. The fit to the data includes the deconvolution with the IRF.}
\label{fig3}
\end{figure*}
Next, the quantum optical properties of the X of three different QDs under resonant $\pi$-pulse excitation were measured in more detail. An overview of the results is given in Table~\ref{table1}, while the corresponding measurements of QD1 are shown in Figure~\ref{fig3}a-d.
First, the decay time was measured in a time-correlated single-photon counting (TCSPC) experiment by correlating the signal of an APD with $\approx$~\SI{60}{\pico \second} time resolution with the laser trigger. A mono-exponential fit including the convolution of the system response (compare Figure~\ref{fig3}a) gives a decay time of \SI{201(1)}{\pico \second}. The absence of a beating in the measurement indicates that either the fine-structure splitting of the X is close to zero or that the two dipoles are exactly parallel and orthogonal to the WG as indicated by the sketch in Figure~\ref{fig3}a. In the latter case, which we believe to be accurate since no beating was observed in any of the TCSPC measurements, only the orthogonal component can couple to the TE-mode of the WG. Note that in order to excite the orthogonal dipole component, the polarization of the excitation laser has to be matching, which may increase the coupling to the TE-mode and thus the amount of laser light in the WG \cite{Makhonin.ea:2014}. However, while a dependence of the laser stray light on the position as well as the wavelength was found, which we ascribe to the presence of local stray centers, no significant increase in laser stray light was found when switching between parallel and orthogonal polarization. In contrast, exciting only one dipole is highly beneficial, as it removes the influence of the fine-structure on the wave-package, which is important for remote two-photon interference experiments. Furthermore, since the laser is suppressed geometrically, driving only one dipole component means no additional losses, which are inherent to excitation based on polarization-suppression \cite{Huber.ea:2020}.\\
\begin{table*}[t!]
  \caption{Overview of state preparation fidelity, decay time $\tau$, linewidth $\Delta f$, and both raw and corrected g$^{(2)}$-values and two-photon interference visibilities (\SI{6.57}{\nano \second} time delay), respectively, measured for three different QDs.}
  \label{table1}
  \begin{tabular}{llllllll}
    \hline
    \multicolumn{1}{c}{QD \#} & \multicolumn{1}{c}{Prep. Fid.} & \multicolumn{1}{c}{$\tau$} & \multicolumn{1}{c}{$\Delta f$} & \multicolumn{1}{c}{g$^{(2)}_{\text{raw}}$(0)} & \multicolumn{1}{c}{g$^{(2)}_{\text{bgc.}}$(0)} &\multicolumn{1}{c}{V$_{\text{TPI, raw}}$} & \multicolumn{1}{c}{V$_{\text{TPI, full-corr.}}$} \\
    \hline
		QD1 & \SI{58.4(25)}{\percent} & \SI{201(1)}{\pico \second} & \SI{4.82(9)}{\giga \hertz} & \num{0.142(3)} & \num{0.078(10)} & \num{0.760(12)} & \num{0.859(15)}\\
    QD2   & \SI{58.9(28)}{\percent} & \SI{135(1)}{\pico \second} & \SI{3.74(8)}{\giga \hertz} & \num{0.209(2)} & \num{0.168(2)} & \num{0.775(14)} & \num{0.939(4)}\\
    QD3  & \SI{63.8(49)}{\percent} & \SI{207(1)}{\pico \second} & \SI{8.79(20)}{\giga \hertz} & \num{0.121(2)} & \num{0.071(9)} & \num{0.714(13)} & \num{0.814(16)}\\
    \hline
  \end{tabular}
\end{table*}
Key for two-photon interference experiments, in particular with remote QDs, is the linewidth of the transition. To classify our emitters in this regard, the linewidths of several Xs under resonant excitation were measured through high-resolution spectroscopy, employing a Fabry-P\'{e}rot interferometer. While the whole statistics is discussed later on, a representative scan for QD1 is plotted in Figure~\ref{fig3}b. In this case, a linewidth of FWHM$_\text{Voigt}$=\SI{4.82(9)}{\giga \hertz} is obtained, which is a factor of $\approx$~6 to the Fourier limit (\SI{0.79}{\giga \hertz}). Since we expect mostly inhomogeneous spectral broadening due to spectral diffusion induced by the nearby WG-sidewalls and residual impurities in the sample, the Lorentzian part of the Voigt fit was fixed to the Fourier-transform limit and only the Gaussian part was fitted. \\
Following the demonstration of resonant excitation without requiring additional off-chip laser-filtering, the single-photon operation of the device needs to be proven. To do so, the g$^{(2)}(\tau)$ is measured using the on-chip MMI as the 50/50 beamsplitter in a Hanbury-Brown and Twiss interferometer. The signal of the two output arms is directly sent to two fiber-coupled APDs with timing resolutions of $\approx$\SI{250}{\pico \second}. The corresponding measurement is shown in Figure~\ref{fig3}c. A strong bunching can be observed, which is well known for DEE GaAs-QDs without stabilization of the electronic environment \cite{Zhai.ea:2020}. In order to correctly extract the Poisson level, the measurement was correlated up to \SI{2}{\milli \second} in a binning corresponding to the repetition rate of the laser (\SI{6.57}{\nano \second}, see inset of Figure~\ref{fig3}c) and was fitted to extract the bunching parameters. From the fit, two different bunching time scales of $\approx$~\SI{65}{\micro \second} and $\approx$~\SI{125}{\micro \second} are obtained, resulting in an optical on-time of \SI{50.8(4)}{\percent}. Similar values were found for all measured QDs and the bunching seemed to be stable between different measurements. After normalization, g$^{(2)}_{\text{raw}}$(0)=\num{0.142(3)} is extracted, by comparing the area of the central peak to the average of 114 side-peaks (corrected for the bunching), with an integration window of \SI{6.57}{\nano \second}. Note that a considerable contribution of the non-vanishing g$^{(2)}$(0) stems from an uncorrelated background, that cannot be explained by the detector dark counts. We ascribe it to a CW-contribution of our laser, since we also see a background when directly measuring the instrumental response function (IRF). Fitting the data including a convolution with the system response allows a correction for the background resulting in g$^{(2)}_{\text{bgc.}}$(0)=\num{0.078(10)}. We ascribe the rest of the non-vanishing g$^{(2)}$(0) to a small amount of pulsed laser stray light, as well as some chance for a re-excitation process to occur within the same pulse.\\
After showing the generation, guiding and on-chip splitting of resonantly excited single photons, the last step consists in benchmarking the two-photon interference visibility of consecutive photons from the same QD. To do so, photons with a time separation of \SI{6.57}{\nano \second} were generated and brought to interfere. The on-chip MMI was again used as a part of the setup, here as first beamsplitter in the Mach-Zehnder interferometer. One of the two detection paths was then delayed by \SI{6.57}{\nano \second} before entering a fiber beamsplitter, where the interference took place. A short free space setup including wave plates for polarization control was implemented before the fiber beamsplitter, which allowed switching between parallel and orthogonal polarization. The corresponding measurements for QD1 are shown in Figure~\ref{fig3}d. Normalization of the data is done similarly to the g$^{(2)}(\tau)$-measurements by a coarse binning to extract the bunching parameters. As mentioned before, no significant changes of the bunching behavior between the different measurements was found. The raw two-photon interference visibility is extracted by comparing the areas of the central peak for the parallel and the orthogonal measurement, with each being normalized to their respective Poisson level, again using an integration window of \SI{6.57}{\nano \second}. Following the formula $\text{V}_\text{TPI}=1-\frac{\text{g}^{(2)}_{\parallel}(0)}{\text{g}^{(2)}_{\perp}(0)}$, a visibility  of V$_{\text{TPI, raw}}$=\num{0.760(12)} is obtained. A fit to the data allows a full correction of the value, including the background counts, the non-perfect g$^{(2)}$(0) and a deviation of the fiber beamsplitter from the ideal 50/50 splitting ratio, resulting in V$_{\text{TPI, full-corr.}}$=\num{0.859(15)}. This indicates that most of the line broadening happens on time scales larger than the delay used in the measurement, as the expected visibility for two remote sources with properties like QD1 would be only 0.30. This value could be further improved by improving the material quality and embedding the QDs in a diode structure \cite{Zhai.ea:2020, Zhai.ea:2022}. The measurements on QD2 and QD3 confirm this behavior, showing high corrected visibilities of \num{0.814(16)} and \num{0.939(4)}, respectively, despite QD3 having more than twice the linewidth of QD2.\\                
Finally, statistical measurements on a large amount of QDs located in different MMIs were performed, beginning with the wavelength distribution of the Xs, as shown in the histogram in Figure~\ref{fig4}a. Averaging over 104 QDs, a central wavelength of \SI{781.71}{\nano \meter} with a standard deviation of \SI{3.53}{\nano \meter} is obtained, underlining the small wavelength distribution of DEE GaAs QDs. Next, the decay times of 25 Xs were measured, with the corresponding histogram plotted in Figure~\ref{fig4}b. The average decay time is \SI{183}{\pico \second} with a standard deviation of \SI{22}{\pico \second} and thus a bit shorter than the $\approx$~\SI{250}{\pico \second} given in Reference~\citenum{Reindl.ea:2019}. A possible reason could be additional non-radiative decay channels due to the nearby WG sidewalls or a slightly weaker confinement of the X due to the lowered Al-content of the barrier material. For the same 25 Xs, the linewidths were also measured and are plotted in the histogram in Figure~\ref{fig4}c. Here, the average linewidth is \SI{6.73}{\giga \hertz} with a standard deviation of \SI{1.77}{\giga \hertz}. As mentioned earlier, we ascribe the main part of the line broadening to spectral diffusion, which may be increased by the nearby WG sidewalls.\\
\begin{figure*}[t!]
\includegraphics{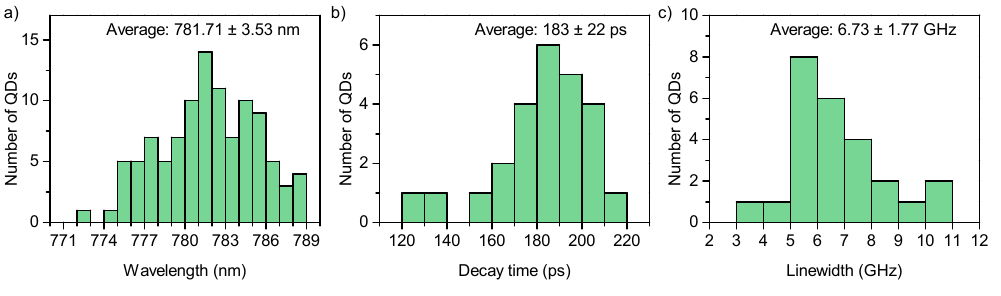}
\caption{Statistical distributions of the Xs, as well as the average and corresponding standard deviations for a) the central wavelength, measured for 104 QDs b) the decay times, measured for 25 QDs c) the linewidth, measured for the same 25 QDs.}
\label{fig4}
\end{figure*}   
In conclusion, we have shown the first integration of DEE GaAs QDs into single-mode WGs, additionally combining them with MMIs to perform an on-chip beamsplitting operation with 50/50 splitting ratio. Both resonant excitation of individual emitters as well as statistical measurements were performed. Three QDs with different linewidths were investigated closer under $\pi$-pulse excitation, showing high two-photon interference visibilities >0.80 and low g$^{(2)}$(0)-values <0.1. Using an excitation laser that is polarized perpendicular to the WGs allowed us to address a single dipole of the X, which is beneficial with respect to the efficiency as well as for potential on-chip two-photon interference experiments from remote sources in the future. When thinking in this direction, the small wavelength distribution (\SI{781.71(353)}{\nano \meter}) and the low average decay time (\SI{183(22)}{\pico \second}) are very promising, although the spectral broadening of the lines (\SI{6.73(177)}{\giga \hertz}) has to be lowered. The latter is mainly attributed to spectral diffusion due to residual impurities in the sample and the nearby WG sidewalls. A surface passivation could help in this regard \cite{Liu.ea:2018_2, Manna.ea:2020}, as well as the integration into diodes \cite{Zhai.ea:2020, Zhai.ea:2022} or deterministic fabrication \cite{Schnauber.ea:2018, Pregnolato.ea:2020} to maximize the distance to the sidewalls. Deterministic fabrication would also be beneficial to preselect QDs that are spatially and spectrally close for integration into the same MMI \cite{Li.ea:2023}. Nevertheless, the integration and resonant excitation of DEE GaAs QDs is an important step towards a high-quality on-chip source of single and indistinguishable photons with a small wavelength distribution and inherently short decay times.   

\subsection{Acknowledgments}
The authors thank Christian Schweikert from the Institute of Electrical and Optical Communications for support with the MMI-simulations and Thomas Reindl and Marion Hagel from the Max Planck Institute for Solid State research for support with the sample preparation. This work was financially supported by the DFG via the project MI 500/32-1, the Austrian Science Fund (FWF) via the project I 4320, the Research Group FG5, the European Union’s Horizon 2020 research and innovation program under Grant Agreements No. 899814 (Qurope) and No. 871130 (Ascent+), the QuantERA II Programme that has received funding from the European Union’s Horizon 2020 research and innovation programme under Grant Agreement No. 101017733 via the project QD-E-QKD and the FFG Grant No. 891366, the Linz Institute of Technology (LIT), the LIT Secure and Correct Systems Lab, supported by the State of Upper Austria.

\subsection{Associated content}
The following files are available free of charge.
\begin{itemize}
  \item Supplementary material: sample growth, simulations of the single-mode WGs and the MMIs, degree of polarization, attenuation measurements, transmission measurements.
\end{itemize}

\subsection{Competing interests}
\noindent The authors declare no competing interests.
\subsection{References}

\begin{thebibliography}{51}%
\makeatletter
\providecommand \@ifxundefined [1]{%
 \@ifx{#1\undefined}
}%
\providecommand \@ifnum [1]{%
 \ifnum #1\expandafter \@firstoftwo
 \else \expandafter \@secondoftwo
 \fi
}%
\providecommand \@ifx [1]{%
 \ifx #1\expandafter \@firstoftwo
 \else \expandafter \@secondoftwo
 \fi
}%
\providecommand \natexlab [1]{#1}%
\providecommand \enquote  [1]{``#1''}%
\providecommand \bibnamefont  [1]{#1}%
\providecommand \bibfnamefont [1]{#1}%
\providecommand \citenamefont [1]{#1}%
\providecommand \href@noop [0]{\@secondoftwo}%
\providecommand \href [0]{\begingroup \@sanitize@url \@href}%
\providecommand \@href[1]{\@@startlink{#1}\@@href}%
\providecommand \@@href[1]{\endgroup#1\@@endlink}%
\providecommand \@sanitize@url [0]{\catcode `\\12\catcode `\$12\catcode
  `\&12\catcode `\#12\catcode `\^12\catcode `\_12\catcode `\%12\relax}%
\providecommand \@@startlink[1]{}%
\providecommand \@@endlink[0]{}%
\providecommand \url  [0]{\begingroup\@sanitize@url \@url }%
\providecommand \@url [1]{\endgroup\@href {#1}{\urlprefix }}%
\providecommand \urlprefix  [0]{URL }%
\providecommand \Eprint [0]{\href }%
\providecommand \doibase [0]{https://doi.org/}%
\providecommand \selectlanguage [0]{\@gobble}%
\providecommand \bibinfo  [0]{\@secondoftwo}%
\providecommand \bibfield  [0]{\@secondoftwo}%
\providecommand \translation [1]{[#1]}%
\providecommand \BibitemOpen [0]{}%
\providecommand \bibitemStop [0]{}%
\providecommand \bibitemNoStop [0]{.\EOS\space}%
\providecommand \EOS [0]{\spacefactor3000\relax}%
\providecommand \BibitemShut  [1]{\csname bibitem#1\endcsname}%
\let\auto@bib@innerbib\@empty
\bibitem [{\citenamefont {Wang}\ \emph {et~al.}(2020)\citenamefont {Wang},
  \citenamefont {Sciarrino}, \citenamefont {Laing},\ and\ \citenamefont
  {Thompson}}]{Wang.ea:2020}%
  \BibitemOpen
  \bibfield  {author} {\bibinfo {author} {\bibfnamefont {J.}~\bibnamefont
  {Wang}}, \bibinfo {author} {\bibfnamefont {F.}~\bibnamefont {Sciarrino}},
  \bibinfo {author} {\bibfnamefont {A.}~\bibnamefont {Laing}},\ and\ \bibinfo
  {author} {\bibfnamefont {M.~G.}\ \bibnamefont {Thompson}},\ }\bibfield
  {title} {\bibinfo {title} {{Integrated photonic quantum technologies}},\
  }\href {https://doi.org/10.1038/s41566-019-0532-1} {\bibfield  {journal}
  {\bibinfo  {journal} {Nature Photonics}\ }\textbf {\bibinfo {volume} {14}},\
  \bibinfo {pages} {273} (\bibinfo {year} {2020})}\BibitemShut {NoStop}%
\bibitem [{\citenamefont {Moody}\ \emph {et~al.}(2022)\citenamefont {Moody},
  \citenamefont {Sorger}, \citenamefont {Blumenthal}, \citenamefont
  {Juodawlkis}, \citenamefont {Loh}, \citenamefont {Sorace-Agaskar},
  \citenamefont {Jones}, \citenamefont {Balram}, \citenamefont {Matthews},
  \citenamefont {Laing}, \citenamefont {Davanco}, \citenamefont {Chang},
  \citenamefont {Bowers}, \citenamefont {Quack}, \citenamefont {Galland},
  \citenamefont {Aharonovich}, \citenamefont {Wolff}, \citenamefont {Schuck},
  \citenamefont {Sinclair}, \citenamefont {Lončar}, \citenamefont
  {Komljenovic}, \citenamefont {Weld}, \citenamefont {Mookherjea},
  \citenamefont {Buckley}, \citenamefont {Radulaski}, \citenamefont
  {Reitzenstein}, \citenamefont {Pingault}, \citenamefont {Machielse},
  \citenamefont {Mukhopadhyay}, \citenamefont {Akimov}, \citenamefont
  {Zheltikov}, \citenamefont {Agarwal}, \citenamefont {Srinivasan},
  \citenamefont {Lu}, \citenamefont {Tang}, \citenamefont {Jiang},
  \citenamefont {McKenna}, \citenamefont {Safavi-Naeini}, \citenamefont
  {Steinhauer}, \citenamefont {Elshaari}, \citenamefont {Zwiller},
  \citenamefont {Davids}, \citenamefont {Martinez}, \citenamefont {Gehl},
  \citenamefont {Chiaverini}, \citenamefont {Mehta}, \citenamefont {Romero},
  \citenamefont {Lingaraju}, \citenamefont {Weiner}, \citenamefont {Peace},
  \citenamefont {Cernansky}, \citenamefont {Lobino}, \citenamefont {Diamanti},
  \citenamefont {Vidarte},\ and\ \citenamefont {Camacho}}]{Moody.ea:2022}%
  \BibitemOpen
  \bibfield  {author} {\bibinfo {author} {\bibfnamefont {G.}~\bibnamefont
  {Moody}}, \bibinfo {author} {\bibfnamefont {V.~J.}\ \bibnamefont {Sorger}},
  \bibinfo {author} {\bibfnamefont {D.~J.}\ \bibnamefont {Blumenthal}},
  \bibinfo {author} {\bibfnamefont {P.~W.}\ \bibnamefont {Juodawlkis}},
  \bibinfo {author} {\bibfnamefont {W.}~\bibnamefont {Loh}}, \bibinfo {author}
  {\bibfnamefont {C.}~\bibnamefont {Sorace-Agaskar}}, \bibinfo {author}
  {\bibfnamefont {A.~E.}\ \bibnamefont {Jones}}, \bibinfo {author}
  {\bibfnamefont {K.~C.}\ \bibnamefont {Balram}}, \bibinfo {author}
  {\bibfnamefont {J.~C.~F.}\ \bibnamefont {Matthews}}, \bibinfo {author}
  {\bibfnamefont {A.}~\bibnamefont {Laing}}, \bibinfo {author} {\bibfnamefont
  {M.}~\bibnamefont {Davanco}}, \bibinfo {author} {\bibfnamefont
  {L.}~\bibnamefont {Chang}}, \bibinfo {author} {\bibfnamefont {J.~E.}\
  \bibnamefont {Bowers}}, \bibinfo {author} {\bibfnamefont {N.}~\bibnamefont
  {Quack}}, \bibinfo {author} {\bibfnamefont {C.}~\bibnamefont {Galland}},
  \bibinfo {author} {\bibfnamefont {I.}~\bibnamefont {Aharonovich}}, \bibinfo
  {author} {\bibfnamefont {M.~A.}\ \bibnamefont {Wolff}}, \bibinfo {author}
  {\bibfnamefont {C.}~\bibnamefont {Schuck}}, \bibinfo {author} {\bibfnamefont
  {N.}~\bibnamefont {Sinclair}}, \bibinfo {author} {\bibfnamefont
  {M.}~\bibnamefont {Lončar}}, \bibinfo {author} {\bibfnamefont
  {T.}~\bibnamefont {Komljenovic}}, \bibinfo {author} {\bibfnamefont
  {D.}~\bibnamefont {Weld}}, \bibinfo {author} {\bibfnamefont {S.}~\bibnamefont
  {Mookherjea}}, \bibinfo {author} {\bibfnamefont {S.}~\bibnamefont {Buckley}},
  \bibinfo {author} {\bibfnamefont {M.}~\bibnamefont {Radulaski}}, \bibinfo
  {author} {\bibfnamefont {S.}~\bibnamefont {Reitzenstein}}, \bibinfo {author}
  {\bibfnamefont {B.}~\bibnamefont {Pingault}}, \bibinfo {author}
  {\bibfnamefont {B.}~\bibnamefont {Machielse}}, \bibinfo {author}
  {\bibfnamefont {D.}~\bibnamefont {Mukhopadhyay}}, \bibinfo {author}
  {\bibfnamefont {A.}~\bibnamefont {Akimov}}, \bibinfo {author} {\bibfnamefont
  {A.}~\bibnamefont {Zheltikov}}, \bibinfo {author} {\bibfnamefont {G.~S.}\
  \bibnamefont {Agarwal}}, \bibinfo {author} {\bibfnamefont {K.}~\bibnamefont
  {Srinivasan}}, \bibinfo {author} {\bibfnamefont {J.}~\bibnamefont {Lu}},
  \bibinfo {author} {\bibfnamefont {H.~X.}\ \bibnamefont {Tang}}, \bibinfo
  {author} {\bibfnamefont {W.}~\bibnamefont {Jiang}}, \bibinfo {author}
  {\bibfnamefont {T.~P.}\ \bibnamefont {McKenna}}, \bibinfo {author}
  {\bibfnamefont {A.~H.}\ \bibnamefont {Safavi-Naeini}}, \bibinfo {author}
  {\bibfnamefont {S.}~\bibnamefont {Steinhauer}}, \bibinfo {author}
  {\bibfnamefont {A.~W.}\ \bibnamefont {Elshaari}}, \bibinfo {author}
  {\bibfnamefont {V.}~\bibnamefont {Zwiller}}, \bibinfo {author} {\bibfnamefont
  {P.~S.}\ \bibnamefont {Davids}}, \bibinfo {author} {\bibfnamefont
  {N.}~\bibnamefont {Martinez}}, \bibinfo {author} {\bibfnamefont
  {M.}~\bibnamefont {Gehl}}, \bibinfo {author} {\bibfnamefont {J.}~\bibnamefont
  {Chiaverini}}, \bibinfo {author} {\bibfnamefont {K.~K.}\ \bibnamefont
  {Mehta}}, \bibinfo {author} {\bibfnamefont {J.}~\bibnamefont {Romero}},
  \bibinfo {author} {\bibfnamefont {N.~B.}\ \bibnamefont {Lingaraju}}, \bibinfo
  {author} {\bibfnamefont {A.~M.}\ \bibnamefont {Weiner}}, \bibinfo {author}
  {\bibfnamefont {D.}~\bibnamefont {Peace}}, \bibinfo {author} {\bibfnamefont
  {R.}~\bibnamefont {Cernansky}}, \bibinfo {author} {\bibfnamefont
  {M.}~\bibnamefont {Lobino}}, \bibinfo {author} {\bibfnamefont
  {E.}~\bibnamefont {Diamanti}}, \bibinfo {author} {\bibfnamefont {L.~T.}\
  \bibnamefont {Vidarte}},\ and\ \bibinfo {author} {\bibfnamefont {R.~M.}\
  \bibnamefont {Camacho}},\ }\bibfield  {title} {\bibinfo {title} {{2022
  Roadmap on integrated quantum photonics}},\ }\href
  {https://doi.org/10.1088/2515-7647/ac1ef4} {\bibfield  {journal} {\bibinfo
  {journal} {Journal of Physics: Photonics}\ }\textbf {\bibinfo {volume} {4}},\
  \bibinfo {pages} {012501} (\bibinfo {year} {2022})}\BibitemShut {NoStop}%
\bibitem [{\citenamefont {Schweickert}\ \emph {et~al.}(2018)\citenamefont
  {Schweickert}, \citenamefont {Jöns}, \citenamefont {Zeuner}, \citenamefont
  {da~Silva}, \citenamefont {Huang}, \citenamefont {Lettner}, \citenamefont
  {Reindl}, \citenamefont {Zichi}, \citenamefont {Trotta}, \citenamefont
  {Rastelli},\ and\ \citenamefont {Zwiller}}]{Schweickert.ea:2018}%
  \BibitemOpen
  \bibfield  {author} {\bibinfo {author} {\bibfnamefont {L.}~\bibnamefont
  {Schweickert}}, \bibinfo {author} {\bibfnamefont {K.~D.}\ \bibnamefont
  {Jöns}}, \bibinfo {author} {\bibfnamefont {K.~D.}\ \bibnamefont {Zeuner}},
  \bibinfo {author} {\bibfnamefont {S.~F.~C.}\ \bibnamefont {da~Silva}},
  \bibinfo {author} {\bibfnamefont {H.}~\bibnamefont {Huang}}, \bibinfo
  {author} {\bibfnamefont {T.}~\bibnamefont {Lettner}}, \bibinfo {author}
  {\bibfnamefont {M.}~\bibnamefont {Reindl}}, \bibinfo {author} {\bibfnamefont
  {J.}~\bibnamefont {Zichi}}, \bibinfo {author} {\bibfnamefont
  {R.}~\bibnamefont {Trotta}}, \bibinfo {author} {\bibfnamefont
  {A.}~\bibnamefont {Rastelli}},\ and\ \bibinfo {author} {\bibfnamefont
  {V.}~\bibnamefont {Zwiller}},\ }\bibfield  {title} {\bibinfo {title}
  {{On-demand generation of background-free single photons from a solid-state
  source}},\ }\href {https://doi.org/10.1063/1.5020038} {\bibfield  {journal}
  {\bibinfo  {journal} {Applied Physics Letters}\ }\textbf {\bibinfo {volume}
  {112}},\ \bibinfo {pages} {093106} (\bibinfo {year} {2018})}\BibitemShut
  {NoStop}%
\bibitem [{\citenamefont {Wang}\ \emph {et~al.}(2019)\citenamefont {Wang},
  \citenamefont {He}, \citenamefont {Chung}, \citenamefont {Hu}, \citenamefont
  {Yu}, \citenamefont {Chen}, \citenamefont {Ding}, \citenamefont {Chen},
  \citenamefont {Qin}, \citenamefont {Yang}, \citenamefont {Liu}, \citenamefont
  {Duan}, \citenamefont {Li}, \citenamefont {Gerhardt}, \citenamefont
  {Winkler}, \citenamefont {Jurkat}, \citenamefont {Wang}, \citenamefont
  {Gregersen}, \citenamefont {Huo}, \citenamefont {Dai}, \citenamefont {Yu},
  \citenamefont {Höfling}, \citenamefont {Lu},\ and\ \citenamefont
  {Pan}}]{Wang.ea:2019}%
  \BibitemOpen
  \bibfield  {author} {\bibinfo {author} {\bibfnamefont {H.}~\bibnamefont
  {Wang}}, \bibinfo {author} {\bibfnamefont {Y.-M.}\ \bibnamefont {He}},
  \bibinfo {author} {\bibfnamefont {T.-H.}\ \bibnamefont {Chung}}, \bibinfo
  {author} {\bibfnamefont {H.}~\bibnamefont {Hu}}, \bibinfo {author}
  {\bibfnamefont {Y.}~\bibnamefont {Yu}}, \bibinfo {author} {\bibfnamefont
  {S.}~\bibnamefont {Chen}}, \bibinfo {author} {\bibfnamefont {X.}~\bibnamefont
  {Ding}}, \bibinfo {author} {\bibfnamefont {M.-C.}\ \bibnamefont {Chen}},
  \bibinfo {author} {\bibfnamefont {J.}~\bibnamefont {Qin}}, \bibinfo {author}
  {\bibfnamefont {X.}~\bibnamefont {Yang}}, \bibinfo {author} {\bibfnamefont
  {R.-Z.}\ \bibnamefont {Liu}}, \bibinfo {author} {\bibfnamefont {Z.-C.}\
  \bibnamefont {Duan}}, \bibinfo {author} {\bibfnamefont {J.-P.}\ \bibnamefont
  {Li}}, \bibinfo {author} {\bibfnamefont {S.}~\bibnamefont {Gerhardt}},
  \bibinfo {author} {\bibfnamefont {K.}~\bibnamefont {Winkler}}, \bibinfo
  {author} {\bibfnamefont {J.}~\bibnamefont {Jurkat}}, \bibinfo {author}
  {\bibfnamefont {L.-J.}\ \bibnamefont {Wang}}, \bibinfo {author}
  {\bibfnamefont {N.}~\bibnamefont {Gregersen}}, \bibinfo {author}
  {\bibfnamefont {Y.-H.}\ \bibnamefont {Huo}}, \bibinfo {author} {\bibfnamefont
  {Q.}~\bibnamefont {Dai}}, \bibinfo {author} {\bibfnamefont {S.}~\bibnamefont
  {Yu}}, \bibinfo {author} {\bibfnamefont {S.}~\bibnamefont {Höfling}},
  \bibinfo {author} {\bibfnamefont {C.-Y.}\ \bibnamefont {Lu}},\ and\ \bibinfo
  {author} {\bibfnamefont {J.-W.}\ \bibnamefont {Pan}},\ }\bibfield  {title}
  {\bibinfo {title} {{Towards optimal single-photon sources from polarized
  microcavities}},\ }\href {https://doi.org/10.1038/s41566-019-0494-3}
  {\bibfield  {journal} {\bibinfo  {journal} {Nature Photonics}\ }\textbf
  {\bibinfo {volume} {13}},\ \bibinfo {pages} {770} (\bibinfo {year}
  {2019})}\BibitemShut {NoStop}%
\bibitem [{\citenamefont {Tomm}\ \emph {et~al.}(2021)\citenamefont {Tomm},
  \citenamefont {Javadi}, \citenamefont {Antoniadis}, \citenamefont {Najer},
  \citenamefont {Löbl}, \citenamefont {Korsch}, \citenamefont {Schott},
  \citenamefont {Valentin}, \citenamefont {Wieck}, \citenamefont {Ludwig},\
  and\ \citenamefont {Warburton}}]{Tomm.ea:2021}%
  \BibitemOpen
  \bibfield  {author} {\bibinfo {author} {\bibfnamefont {N.}~\bibnamefont
  {Tomm}}, \bibinfo {author} {\bibfnamefont {A.}~\bibnamefont {Javadi}},
  \bibinfo {author} {\bibfnamefont {N.~O.}\ \bibnamefont {Antoniadis}},
  \bibinfo {author} {\bibfnamefont {D.}~\bibnamefont {Najer}}, \bibinfo
  {author} {\bibfnamefont {M.~C.}\ \bibnamefont {Löbl}}, \bibinfo {author}
  {\bibfnamefont {A.~R.}\ \bibnamefont {Korsch}}, \bibinfo {author}
  {\bibfnamefont {R.}~\bibnamefont {Schott}}, \bibinfo {author} {\bibfnamefont
  {S.~R.}\ \bibnamefont {Valentin}}, \bibinfo {author} {\bibfnamefont {A.~D.}\
  \bibnamefont {Wieck}}, \bibinfo {author} {\bibfnamefont {A.}~\bibnamefont
  {Ludwig}},\ and\ \bibinfo {author} {\bibfnamefont {R.~J.}\ \bibnamefont
  {Warburton}},\ }\bibfield  {title} {\bibinfo {title} {{A bright and fast
  source of coherent single photons}},\ }\href
  {https://doi.org/10.1038/s41565-020-00831-x} {\bibfield  {journal} {\bibinfo
  {journal} {Nature Nanotechnology}\ }\textbf {\bibinfo {volume} {16}},\
  \bibinfo {pages} {399} (\bibinfo {year} {2021})}\BibitemShut {NoStop}%
\bibitem [{\citenamefont {Nawrath}\ \emph {et~al.}(2023)\citenamefont
  {Nawrath}, \citenamefont {Joos}, \citenamefont {Kolatschek}, \citenamefont
  {Bauer}, \citenamefont {Pruy}, \citenamefont {Hornung}, \citenamefont
  {Fischer}, \citenamefont {Huang}, \citenamefont {Vijayan}, \citenamefont
  {Sittig}, \citenamefont {Jetter}, \citenamefont {Portalupi},\ and\
  \citenamefont {Michler}}]{Nawrath.ea:2023}%
  \BibitemOpen
  \bibfield  {author} {\bibinfo {author} {\bibfnamefont {C.}~\bibnamefont
  {Nawrath}}, \bibinfo {author} {\bibfnamefont {R.}~\bibnamefont {Joos}},
  \bibinfo {author} {\bibfnamefont {S.}~\bibnamefont {Kolatschek}}, \bibinfo
  {author} {\bibfnamefont {S.}~\bibnamefont {Bauer}}, \bibinfo {author}
  {\bibfnamefont {P.}~\bibnamefont {Pruy}}, \bibinfo {author} {\bibfnamefont
  {F.}~\bibnamefont {Hornung}}, \bibinfo {author} {\bibfnamefont
  {J.}~\bibnamefont {Fischer}}, \bibinfo {author} {\bibfnamefont
  {J.}~\bibnamefont {Huang}}, \bibinfo {author} {\bibfnamefont
  {P.}~\bibnamefont {Vijayan}}, \bibinfo {author} {\bibfnamefont
  {R.}~\bibnamefont {Sittig}}, \bibinfo {author} {\bibfnamefont
  {M.}~\bibnamefont {Jetter}}, \bibinfo {author} {\bibfnamefont {S.~L.}\
  \bibnamefont {Portalupi}},\ and\ \bibinfo {author} {\bibfnamefont
  {P.}~\bibnamefont {Michler}},\ }\bibfield  {title} {\bibinfo {title} {{Bright
  Source of Purcell‐Enhanced, Triggered, Single Photons in the Telecom
  C‐Band}},\ }\href {https://doi.org/10.1002/qute.202300111} {\bibfield
  {journal} {\bibinfo  {journal} {Advanced Quantum Technologies}\ ,\ \bibinfo
  {pages} {2300111}} (\bibinfo {year} {2023})}\BibitemShut {NoStop}%
\bibitem [{\citenamefont {Kuhlmann}\ \emph {et~al.}(2015)\citenamefont
  {Kuhlmann}, \citenamefont {Prechtel}, \citenamefont {Houel}, \citenamefont
  {Ludwig}, \citenamefont {Reuter}, \citenamefont {Wieck},\ and\ \citenamefont
  {Warburton}}]{Kuhlmann.ea:2015}%
  \BibitemOpen
  \bibfield  {author} {\bibinfo {author} {\bibfnamefont {A.~V.}\ \bibnamefont
  {Kuhlmann}}, \bibinfo {author} {\bibfnamefont {J.~H.}\ \bibnamefont
  {Prechtel}}, \bibinfo {author} {\bibfnamefont {J.}~\bibnamefont {Houel}},
  \bibinfo {author} {\bibfnamefont {A.}~\bibnamefont {Ludwig}}, \bibinfo
  {author} {\bibfnamefont {D.}~\bibnamefont {Reuter}}, \bibinfo {author}
  {\bibfnamefont {A.~D.}\ \bibnamefont {Wieck}},\ and\ \bibinfo {author}
  {\bibfnamefont {R.~J.}\ \bibnamefont {Warburton}},\ }\bibfield  {title}
  {\bibinfo {title} {{Transform-limited single photons from a single quantum
  dot}},\ }\href {https://doi.org/10.1038/ncomms9204} {\bibfield  {journal}
  {\bibinfo  {journal} {Nature Communications}\ }\textbf {\bibinfo {volume}
  {6}},\ \bibinfo {pages} {8204} (\bibinfo {year} {2015})}\BibitemShut
  {NoStop}%
\bibitem [{\citenamefont {Pedersen}\ \emph {et~al.}(2020)\citenamefont
  {Pedersen}, \citenamefont {Wang}, \citenamefont {Olesen}, \citenamefont
  {Scholz}, \citenamefont {Wieck}, \citenamefont {Ludwig}, \citenamefont
  {Löbl}, \citenamefont {Warburton}, \citenamefont {Midolo}, \citenamefont
  {Uppu},\ and\ \citenamefont {Lodahl}}]{Pedersen.ea:2020}%
  \BibitemOpen
  \bibfield  {author} {\bibinfo {author} {\bibfnamefont {F.~T.}\ \bibnamefont
  {Pedersen}}, \bibinfo {author} {\bibfnamefont {Y.}~\bibnamefont {Wang}},
  \bibinfo {author} {\bibfnamefont {C.~T.}\ \bibnamefont {Olesen}}, \bibinfo
  {author} {\bibfnamefont {S.}~\bibnamefont {Scholz}}, \bibinfo {author}
  {\bibfnamefont {A.~D.}\ \bibnamefont {Wieck}}, \bibinfo {author}
  {\bibfnamefont {A.}~\bibnamefont {Ludwig}}, \bibinfo {author} {\bibfnamefont
  {M.~C.}\ \bibnamefont {Löbl}}, \bibinfo {author} {\bibfnamefont {R.~J.}\
  \bibnamefont {Warburton}}, \bibinfo {author} {\bibfnamefont {L.}~\bibnamefont
  {Midolo}}, \bibinfo {author} {\bibfnamefont {R.}~\bibnamefont {Uppu}},\ and\
  \bibinfo {author} {\bibfnamefont {P.}~\bibnamefont {Lodahl}},\ }\bibfield
  {title} {\bibinfo {title} {{Near Transform-Limited Quantum Dot Linewidths in
  a Broadband Photonic Crystal Waveguide}},\ }\href
  {https://doi.org/10.1021/acsphotonics.0c00758} {\bibfield  {journal}
  {\bibinfo  {journal} {ACS Photonics}\ }\textbf {\bibinfo {volume} {7}},\
  \bibinfo {pages} {2343} (\bibinfo {year} {2020})}\BibitemShut {NoStop}%
\bibitem [{\citenamefont {Strobel}\ \emph {et~al.}(2023)\citenamefont
  {Strobel}, \citenamefont {Weber}, \citenamefont {Schmidt}, \citenamefont
  {Wagner}, \citenamefont {Engel}, \citenamefont {Jetter}, \citenamefont
  {Wieck}, \citenamefont {Portalupi}, \citenamefont {Ludwig},\ and\
  \citenamefont {Michler}}]{Strobel.ea:2023}%
  \BibitemOpen
  \bibfield  {author} {\bibinfo {author} {\bibfnamefont {T.}~\bibnamefont
  {Strobel}}, \bibinfo {author} {\bibfnamefont {J.~H.}\ \bibnamefont {Weber}},
  \bibinfo {author} {\bibfnamefont {M.}~\bibnamefont {Schmidt}}, \bibinfo
  {author} {\bibfnamefont {L.}~\bibnamefont {Wagner}}, \bibinfo {author}
  {\bibfnamefont {L.}~\bibnamefont {Engel}}, \bibinfo {author} {\bibfnamefont
  {M.}~\bibnamefont {Jetter}}, \bibinfo {author} {\bibfnamefont {A.~D.}\
  \bibnamefont {Wieck}}, \bibinfo {author} {\bibfnamefont {S.~L.}\ \bibnamefont
  {Portalupi}}, \bibinfo {author} {\bibfnamefont {A.}~\bibnamefont {Ludwig}},\
  and\ \bibinfo {author} {\bibfnamefont {P.}~\bibnamefont {Michler}},\
  }\bibfield  {title} {\bibinfo {title} {{A Unipolar Quantum Dot Diode
  Structure for Advanced Quantum Light Sources}},\ }\href
  {https://doi.org/10.1021/acs.nanolett.3c01658} {\bibfield  {journal}
  {\bibinfo  {journal} {Nano Letters}\ }\textbf {\bibinfo {volume} {23}},\
  \bibinfo {pages} {6574} (\bibinfo {year} {2023})}\BibitemShut {NoStop}%
\bibitem [{\citenamefont {Schwagmann}\ \emph {et~al.}(2011)\citenamefont
  {Schwagmann}, \citenamefont {Kalliakos}, \citenamefont {Farrer},
  \citenamefont {Griffiths}, \citenamefont {Jones}, \citenamefont {Ritchie},\
  and\ \citenamefont {Shields}}]{Schwagmann.ea:2011}%
  \BibitemOpen
  \bibfield  {author} {\bibinfo {author} {\bibfnamefont {A.}~\bibnamefont
  {Schwagmann}}, \bibinfo {author} {\bibfnamefont {S.}~\bibnamefont
  {Kalliakos}}, \bibinfo {author} {\bibfnamefont {I.}~\bibnamefont {Farrer}},
  \bibinfo {author} {\bibfnamefont {J.~P.}\ \bibnamefont {Griffiths}}, \bibinfo
  {author} {\bibfnamefont {G.~A.~C.}\ \bibnamefont {Jones}}, \bibinfo {author}
  {\bibfnamefont {D.~A.}\ \bibnamefont {Ritchie}},\ and\ \bibinfo {author}
  {\bibfnamefont {A.~J.}\ \bibnamefont {Shields}},\ }\bibfield  {title}
  {\bibinfo {title} {{On-chip single photon emission from an integrated
  semiconductor quantum dot into a photonic crystal waveguide}},\ }\href
  {https://doi.org/10.1063/1.3672214} {\bibfield  {journal} {\bibinfo
  {journal} {Applied Physics Letters}\ }\textbf {\bibinfo {volume} {99}},\
  \bibinfo {pages} {261108} (\bibinfo {year} {2011})}\BibitemShut {NoStop}%
\bibitem [{\citenamefont {Laucht}\ \emph {et~al.}(2012)\citenamefont {Laucht},
  \citenamefont {Pütz}, \citenamefont {Günthner}, \citenamefont {Hauke},
  \citenamefont {Saive}, \citenamefont {Frédérick}, \citenamefont {Bichler},
  \citenamefont {Amann}, \citenamefont {Holleitner}, \citenamefont {Kaniber},\
  and\ \citenamefont {Finley}}]{Laucht.ea:2012}%
  \BibitemOpen
  \bibfield  {author} {\bibinfo {author} {\bibfnamefont {A.}~\bibnamefont
  {Laucht}}, \bibinfo {author} {\bibfnamefont {S.}~\bibnamefont {Pütz}},
  \bibinfo {author} {\bibfnamefont {T.}~\bibnamefont {Günthner}}, \bibinfo
  {author} {\bibfnamefont {N.}~\bibnamefont {Hauke}}, \bibinfo {author}
  {\bibfnamefont {R.}~\bibnamefont {Saive}}, \bibinfo {author} {\bibfnamefont
  {S.}~\bibnamefont {Frédérick}}, \bibinfo {author} {\bibfnamefont
  {M.}~\bibnamefont {Bichler}}, \bibinfo {author} {\bibfnamefont {M.-C.}\
  \bibnamefont {Amann}}, \bibinfo {author} {\bibfnamefont {A.~W.}\ \bibnamefont
  {Holleitner}}, \bibinfo {author} {\bibfnamefont {M.}~\bibnamefont
  {Kaniber}},\ and\ \bibinfo {author} {\bibfnamefont {J.~J.}\ \bibnamefont
  {Finley}},\ }\bibfield  {title} {\bibinfo {title} {{A Waveguide-Coupled
  On-Chip Single-Photon Source}},\ }\href
  {https://doi.org/10.1103/PhysRevX.2.011014} {\bibfield  {journal} {\bibinfo
  {journal} {Physical Review X}\ }\textbf {\bibinfo {volume} {2}},\ \bibinfo
  {pages} {011014} (\bibinfo {year} {2012})}\BibitemShut {NoStop}%
\bibitem [{\citenamefont {Prtljaga}\ \emph {et~al.}(2014)\citenamefont
  {Prtljaga}, \citenamefont {Coles}, \citenamefont {O'Hara}, \citenamefont
  {Royall}, \citenamefont {Clarke}, \citenamefont {Fox},\ and\ \citenamefont
  {Skolnick}}]{Prtljaga.ea:2014}%
  \BibitemOpen
  \bibfield  {author} {\bibinfo {author} {\bibfnamefont {N.}~\bibnamefont
  {Prtljaga}}, \bibinfo {author} {\bibfnamefont {R.~J.}\ \bibnamefont {Coles}},
  \bibinfo {author} {\bibfnamefont {J.}~\bibnamefont {O'Hara}}, \bibinfo
  {author} {\bibfnamefont {B.}~\bibnamefont {Royall}}, \bibinfo {author}
  {\bibfnamefont {E.}~\bibnamefont {Clarke}}, \bibinfo {author} {\bibfnamefont
  {A.~M.}\ \bibnamefont {Fox}},\ and\ \bibinfo {author} {\bibfnamefont {M.~S.}\
  \bibnamefont {Skolnick}},\ }\bibfield  {title} {\bibinfo {title} {{Monolithic
  integration of a quantum emitter with a compact on-chip beam-splitter}},\
  }\href {https://doi.org/10.1063/1.4883374} {\bibfield  {journal} {\bibinfo
  {journal} {Applied Physics Letters}\ }\textbf {\bibinfo {volume} {104}},\
  \bibinfo {pages} {231107} (\bibinfo {year} {2014})}\BibitemShut {NoStop}%
\bibitem [{\citenamefont {Kiršanskė}\ \emph {et~al.}(2017)\citenamefont
  {Kiršanskė}, \citenamefont {Thyrrestrup}, \citenamefont {Daveau},
  \citenamefont {Dreeßen}, \citenamefont {Pregnolato}, \citenamefont {Midolo},
  \citenamefont {Tighineanu}, \citenamefont {Javadi}, \citenamefont {Stobbe},
  \citenamefont {Schott}, \citenamefont {Ludwig}, \citenamefont {Wieck},
  \citenamefont {Park}, \citenamefont {Song}, \citenamefont {Kuhlmann},
  \citenamefont {Söllner}, \citenamefont {Löbl}, \citenamefont {Warburton},\
  and\ \citenamefont {Lodahl}}]{Kirsanske.ea:2017}%
  \BibitemOpen
  \bibfield  {author} {\bibinfo {author} {\bibfnamefont {G.}~\bibnamefont
  {Kiršanskė}}, \bibinfo {author} {\bibfnamefont {H.}~\bibnamefont
  {Thyrrestrup}}, \bibinfo {author} {\bibfnamefont {R.~S.}\ \bibnamefont
  {Daveau}}, \bibinfo {author} {\bibfnamefont {C.~L.}\ \bibnamefont
  {Dreeßen}}, \bibinfo {author} {\bibfnamefont {T.}~\bibnamefont
  {Pregnolato}}, \bibinfo {author} {\bibfnamefont {L.}~\bibnamefont {Midolo}},
  \bibinfo {author} {\bibfnamefont {P.}~\bibnamefont {Tighineanu}}, \bibinfo
  {author} {\bibfnamefont {A.}~\bibnamefont {Javadi}}, \bibinfo {author}
  {\bibfnamefont {S.}~\bibnamefont {Stobbe}}, \bibinfo {author} {\bibfnamefont
  {R.}~\bibnamefont {Schott}}, \bibinfo {author} {\bibfnamefont
  {A.}~\bibnamefont {Ludwig}}, \bibinfo {author} {\bibfnamefont {A.~D.}\
  \bibnamefont {Wieck}}, \bibinfo {author} {\bibfnamefont {S.~I.}\ \bibnamefont
  {Park}}, \bibinfo {author} {\bibfnamefont {J.~D.}\ \bibnamefont {Song}},
  \bibinfo {author} {\bibfnamefont {A.~V.}\ \bibnamefont {Kuhlmann}}, \bibinfo
  {author} {\bibfnamefont {I.}~\bibnamefont {Söllner}}, \bibinfo {author}
  {\bibfnamefont {M.~C.}\ \bibnamefont {Löbl}}, \bibinfo {author}
  {\bibfnamefont {R.~J.}\ \bibnamefont {Warburton}},\ and\ \bibinfo {author}
  {\bibfnamefont {P.}~\bibnamefont {Lodahl}},\ }\bibfield  {title} {\bibinfo
  {title} {{Indistinguishable and efficient single photons from a quantum dot
  in a planar nanobeam waveguide}},\ }\href
  {https://doi.org/10.1103/PhysRevB.96.165306} {\bibfield  {journal} {\bibinfo
  {journal} {Physical Review B}\ }\textbf {\bibinfo {volume} {96}},\ \bibinfo
  {pages} {165306} (\bibinfo {year} {2017})}\BibitemShut {NoStop}%
\bibitem [{\citenamefont {Schnauber}\ \emph {et~al.}(2018)\citenamefont
  {Schnauber}, \citenamefont {Schall}, \citenamefont {Bounouar}, \citenamefont
  {Höhne}, \citenamefont {Park}, \citenamefont {Ryu}, \citenamefont {Heindel},
  \citenamefont {Burger}, \citenamefont {Song}, \citenamefont {Rodt},\ and\
  \citenamefont {Reitzenstein}}]{Schnauber.ea:2018}%
  \BibitemOpen
  \bibfield  {author} {\bibinfo {author} {\bibfnamefont {P.}~\bibnamefont
  {Schnauber}}, \bibinfo {author} {\bibfnamefont {J.}~\bibnamefont {Schall}},
  \bibinfo {author} {\bibfnamefont {S.}~\bibnamefont {Bounouar}}, \bibinfo
  {author} {\bibfnamefont {T.}~\bibnamefont {Höhne}}, \bibinfo {author}
  {\bibfnamefont {S.-I.}\ \bibnamefont {Park}}, \bibinfo {author}
  {\bibfnamefont {G.-H.}\ \bibnamefont {Ryu}}, \bibinfo {author} {\bibfnamefont
  {T.}~\bibnamefont {Heindel}}, \bibinfo {author} {\bibfnamefont
  {S.}~\bibnamefont {Burger}}, \bibinfo {author} {\bibfnamefont {J.-D.}\
  \bibnamefont {Song}}, \bibinfo {author} {\bibfnamefont {S.}~\bibnamefont
  {Rodt}},\ and\ \bibinfo {author} {\bibfnamefont {S.}~\bibnamefont
  {Reitzenstein}},\ }\bibfield  {title} {\bibinfo {title} {{Deterministic
  Integration of Quantum Dots into on-Chip Multimode Interference Beamsplitters
  Using in Situ Electron Beam Lithography}},\ }\href
  {https://doi.org/10.1021/acs.nanolett.7b05218} {\bibfield  {journal}
  {\bibinfo  {journal} {Nano Letters}\ }\textbf {\bibinfo {volume} {18}},\
  \bibinfo {pages} {2336} (\bibinfo {year} {2018})}\BibitemShut {NoStop}%
\bibitem [{\citenamefont {Schwartz}\ \emph {et~al.}(2018)\citenamefont
  {Schwartz}, \citenamefont {Schmidt}, \citenamefont {Rengstl}, \citenamefont
  {Hornung}, \citenamefont {Hepp}, \citenamefont {Portalupi}, \citenamefont
  {llin}, \citenamefont {Jetter}, \citenamefont {Siegel},\ and\ \citenamefont
  {Michler}}]{Schwartz.ea:2018}%
  \BibitemOpen
  \bibfield  {author} {\bibinfo {author} {\bibfnamefont {M.}~\bibnamefont
  {Schwartz}}, \bibinfo {author} {\bibfnamefont {E.}~\bibnamefont {Schmidt}},
  \bibinfo {author} {\bibfnamefont {U.}~\bibnamefont {Rengstl}}, \bibinfo
  {author} {\bibfnamefont {F.}~\bibnamefont {Hornung}}, \bibinfo {author}
  {\bibfnamefont {S.}~\bibnamefont {Hepp}}, \bibinfo {author} {\bibfnamefont
  {S.~L.}\ \bibnamefont {Portalupi}}, \bibinfo {author} {\bibfnamefont
  {K.}~\bibnamefont {llin}}, \bibinfo {author} {\bibfnamefont {M.}~\bibnamefont
  {Jetter}}, \bibinfo {author} {\bibfnamefont {M.}~\bibnamefont {Siegel}},\
  and\ \bibinfo {author} {\bibfnamefont {P.}~\bibnamefont {Michler}},\
  }\bibfield  {title} {\bibinfo {title} {{Fully On-Chip Single-Photon
  Hanbury-Brown and Twiss Experiment on a Monolithic
  Semiconductor–Superconductor Platform}},\ }\href
  {https://doi.org/10.1021/acs.nanolett.8b02794} {\bibfield  {journal}
  {\bibinfo  {journal} {Nano Letters}\ }\textbf {\bibinfo {volume} {18}},\
  \bibinfo {pages} {6892} (\bibinfo {year} {2018})}\BibitemShut {NoStop}%
\bibitem [{\citenamefont {Zadeh}\ \emph {et~al.}(2016)\citenamefont {Zadeh},
  \citenamefont {Elshaari}, \citenamefont {Jöns}, \citenamefont {Fognini},
  \citenamefont {Dalacu}, \citenamefont {Poole}, \citenamefont {Reimer},\ and\
  \citenamefont {Zwiller}}]{Zadeh.ea:2016}%
  \BibitemOpen
  \bibfield  {author} {\bibinfo {author} {\bibfnamefont {I.~E.}\ \bibnamefont
  {Zadeh}}, \bibinfo {author} {\bibfnamefont {A.~W.}\ \bibnamefont {Elshaari}},
  \bibinfo {author} {\bibfnamefont {K.~D.}\ \bibnamefont {Jöns}}, \bibinfo
  {author} {\bibfnamefont {A.}~\bibnamefont {Fognini}}, \bibinfo {author}
  {\bibfnamefont {D.}~\bibnamefont {Dalacu}}, \bibinfo {author} {\bibfnamefont
  {P.~J.}\ \bibnamefont {Poole}}, \bibinfo {author} {\bibfnamefont {M.~E.}\
  \bibnamefont {Reimer}},\ and\ \bibinfo {author} {\bibfnamefont
  {V.}~\bibnamefont {Zwiller}},\ }\bibfield  {title} {\bibinfo {title}
  {{Deterministic Integration of Single Photon Sources in Silicon Based
  Photonic Circuits}},\ }\href {https://doi.org/10.1021/acs.nanolett.5b04709}
  {\bibfield  {journal} {\bibinfo  {journal} {Nano Letters}\ }\textbf {\bibinfo
  {volume} {16}},\ \bibinfo {pages} {2289} (\bibinfo {year}
  {2016})}\BibitemShut {NoStop}%
\bibitem [{\citenamefont {Davanco}\ \emph {et~al.}(2017)\citenamefont
  {Davanco}, \citenamefont {Liu}, \citenamefont {Sapienza}, \citenamefont
  {Zhang}, \citenamefont {Cardoso}, \citenamefont {Verma}, \citenamefont
  {Mirin}, \citenamefont {Nam}, \citenamefont {Liu},\ and\ \citenamefont
  {Srinivasan}}]{Davanco.ea:2017}%
  \BibitemOpen
  \bibfield  {author} {\bibinfo {author} {\bibfnamefont {M.}~\bibnamefont
  {Davanco}}, \bibinfo {author} {\bibfnamefont {J.}~\bibnamefont {Liu}},
  \bibinfo {author} {\bibfnamefont {L.}~\bibnamefont {Sapienza}}, \bibinfo
  {author} {\bibfnamefont {C.-Z.}\ \bibnamefont {Zhang}}, \bibinfo {author}
  {\bibfnamefont {J.~V. D.~M.}\ \bibnamefont {Cardoso}}, \bibinfo {author}
  {\bibfnamefont {V.}~\bibnamefont {Verma}}, \bibinfo {author} {\bibfnamefont
  {R.}~\bibnamefont {Mirin}}, \bibinfo {author} {\bibfnamefont {S.~W.}\
  \bibnamefont {Nam}}, \bibinfo {author} {\bibfnamefont {L.}~\bibnamefont
  {Liu}},\ and\ \bibinfo {author} {\bibfnamefont {K.}~\bibnamefont
  {Srinivasan}},\ }\bibfield  {title} {\bibinfo {title} {{Heterogeneous
  integration for on-chip quantum photonic circuits with single quantum dot
  devices}},\ }\href {https://doi.org/10.1038/s41467-017-00987-6} {\bibfield
  {journal} {\bibinfo  {journal} {Nature Communications}\ }\textbf {\bibinfo
  {volume} {8}},\ \bibinfo {pages} {889} (\bibinfo {year} {2017})}\BibitemShut
  {NoStop}%
\bibitem [{\citenamefont {Kim}\ \emph {et~al.}(2017)\citenamefont {Kim},
  \citenamefont {Aghaeimeibodi}, \citenamefont {Richardson}, \citenamefont
  {Leavitt}, \citenamefont {Englund},\ and\ \citenamefont
  {Waks}}]{Kim.ea:2017}%
  \BibitemOpen
  \bibfield  {author} {\bibinfo {author} {\bibfnamefont {J.-H.}\ \bibnamefont
  {Kim}}, \bibinfo {author} {\bibfnamefont {S.}~\bibnamefont {Aghaeimeibodi}},
  \bibinfo {author} {\bibfnamefont {C.~J.~K.}\ \bibnamefont {Richardson}},
  \bibinfo {author} {\bibfnamefont {R.~P.}\ \bibnamefont {Leavitt}}, \bibinfo
  {author} {\bibfnamefont {D.}~\bibnamefont {Englund}},\ and\ \bibinfo {author}
  {\bibfnamefont {E.}~\bibnamefont {Waks}},\ }\bibfield  {title} {\bibinfo
  {title} {{Hybrid Integration of Solid-State Quantum Emitters on a Silicon
  Photonic Chip}},\ }\href {https://doi.org/10.1021/acs.nanolett.7b03220}
  {\bibfield  {journal} {\bibinfo  {journal} {Nano Letters}\ }\textbf {\bibinfo
  {volume} {17}},\ \bibinfo {pages} {7394} (\bibinfo {year}
  {2017})}\BibitemShut {NoStop}%
\bibitem [{\citenamefont {Katsumi}\ \emph {et~al.}(2018)\citenamefont
  {Katsumi}, \citenamefont {Ota}, \citenamefont {Kakuda}, \citenamefont
  {Iwamoto},\ and\ \citenamefont {Arakawa}}]{Katsumi.ea:2018}%
  \BibitemOpen
  \bibfield  {author} {\bibinfo {author} {\bibfnamefont {R.}~\bibnamefont
  {Katsumi}}, \bibinfo {author} {\bibfnamefont {Y.}~\bibnamefont {Ota}},
  \bibinfo {author} {\bibfnamefont {M.}~\bibnamefont {Kakuda}}, \bibinfo
  {author} {\bibfnamefont {S.}~\bibnamefont {Iwamoto}},\ and\ \bibinfo {author}
  {\bibfnamefont {Y.}~\bibnamefont {Arakawa}},\ }\bibfield  {title} {\bibinfo
  {title} {{Transfer-printed single-photon sources coupled to wire
  waveguides}},\ }\href {https://doi.org/10.1364/OPTICA.5.000691} {\bibfield
  {journal} {\bibinfo  {journal} {Optica}\ }\textbf {\bibinfo {volume} {5}},\
  \bibinfo {pages} {691} (\bibinfo {year} {2018})}\BibitemShut {NoStop}%
\bibitem [{\citenamefont {Ellis}\ \emph {et~al.}(2018)\citenamefont {Ellis},
  \citenamefont {Bennett}, \citenamefont {Dangel}, \citenamefont {Lee},
  \citenamefont {Griffiths}, \citenamefont {Mitchell}, \citenamefont {Paraiso},
  \citenamefont {Spencer}, \citenamefont {Ritchie},\ and\ \citenamefont
  {Shields}}]{Ellis.ea:2018}%
  \BibitemOpen
  \bibfield  {author} {\bibinfo {author} {\bibfnamefont {D.~J.~P.}\
  \bibnamefont {Ellis}}, \bibinfo {author} {\bibfnamefont {A.~J.}\ \bibnamefont
  {Bennett}}, \bibinfo {author} {\bibfnamefont {C.}~\bibnamefont {Dangel}},
  \bibinfo {author} {\bibfnamefont {J.~P.}\ \bibnamefont {Lee}}, \bibinfo
  {author} {\bibfnamefont {J.~P.}\ \bibnamefont {Griffiths}}, \bibinfo {author}
  {\bibfnamefont {T.~A.}\ \bibnamefont {Mitchell}}, \bibinfo {author}
  {\bibfnamefont {T.-K.}\ \bibnamefont {Paraiso}}, \bibinfo {author}
  {\bibfnamefont {P.}~\bibnamefont {Spencer}}, \bibinfo {author} {\bibfnamefont
  {D.~A.}\ \bibnamefont {Ritchie}},\ and\ \bibinfo {author} {\bibfnamefont
  {A.~J.}\ \bibnamefont {Shields}},\ }\bibfield  {title} {\bibinfo {title}
  {{Independent indistinguishable quantum light sources on a reconfigurable
  photonic integrated circuit}},\ }\href {https://doi.org/10.1063/1.5028339}
  {\bibfield  {journal} {\bibinfo  {journal} {Applied Physics Letters}\
  }\textbf {\bibinfo {volume} {112}},\ \bibinfo {pages} {211104} (\bibinfo
  {year} {2018})}\BibitemShut {NoStop}%
\bibitem [{\citenamefont {Singh}\ \emph {et~al.}(2019)\citenamefont {Singh},
  \citenamefont {Li}, \citenamefont {Liu}, \citenamefont {Yu}, \citenamefont
  {Lu}, \citenamefont {Schneider}, \citenamefont {Höfling}, \citenamefont
  {Lawall}, \citenamefont {Verma}, \citenamefont {Mirin}, \citenamefont {Nam},
  \citenamefont {Liu},\ and\ \citenamefont {Srinivasan}}]{Singh.ea:2018}%
  \BibitemOpen
  \bibfield  {author} {\bibinfo {author} {\bibfnamefont {A.}~\bibnamefont
  {Singh}}, \bibinfo {author} {\bibfnamefont {Q.}~\bibnamefont {Li}}, \bibinfo
  {author} {\bibfnamefont {S.}~\bibnamefont {Liu}}, \bibinfo {author}
  {\bibfnamefont {Y.}~\bibnamefont {Yu}}, \bibinfo {author} {\bibfnamefont
  {X.}~\bibnamefont {Lu}}, \bibinfo {author} {\bibfnamefont {C.}~\bibnamefont
  {Schneider}}, \bibinfo {author} {\bibfnamefont {S.}~\bibnamefont {Höfling}},
  \bibinfo {author} {\bibfnamefont {J.}~\bibnamefont {Lawall}}, \bibinfo
  {author} {\bibfnamefont {V.}~\bibnamefont {Verma}}, \bibinfo {author}
  {\bibfnamefont {R.}~\bibnamefont {Mirin}}, \bibinfo {author} {\bibfnamefont
  {S.~W.}\ \bibnamefont {Nam}}, \bibinfo {author} {\bibfnamefont
  {J.}~\bibnamefont {Liu}},\ and\ \bibinfo {author} {\bibfnamefont
  {K.}~\bibnamefont {Srinivasan}},\ }\bibfield  {title} {\bibinfo {title}
  {{Quantum frequency conversion of a quantum dot single-photon source on a
  nanophotonic chip}},\ }\href {https://doi.org/10.1364/OPTICA.6.000563}
  {\bibfield  {journal} {\bibinfo  {journal} {Optica}\ }\textbf {\bibinfo
  {volume} {6}},\ \bibinfo {pages} {563} (\bibinfo {year} {2019})}\BibitemShut
  {NoStop}%
\bibitem [{\citenamefont {Bauer}\ \emph {et~al.}(2021)\citenamefont {Bauer},
  \citenamefont {Wang}, \citenamefont {Hoppe}, \citenamefont {Nawrath},
  \citenamefont {Fischer}, \citenamefont {Witz}, \citenamefont {Kaschel},
  \citenamefont {Schweikert}, \citenamefont {Jetter}, \citenamefont
  {Portalupi}, \citenamefont {Berroth},\ and\ \citenamefont
  {Michler}}]{Bauer.ea:2021}%
  \BibitemOpen
  \bibfield  {author} {\bibinfo {author} {\bibfnamefont {S.}~\bibnamefont
  {Bauer}}, \bibinfo {author} {\bibfnamefont {D.}~\bibnamefont {Wang}},
  \bibinfo {author} {\bibfnamefont {N.}~\bibnamefont {Hoppe}}, \bibinfo
  {author} {\bibfnamefont {C.}~\bibnamefont {Nawrath}}, \bibinfo {author}
  {\bibfnamefont {J.}~\bibnamefont {Fischer}}, \bibinfo {author} {\bibfnamefont
  {N.}~\bibnamefont {Witz}}, \bibinfo {author} {\bibfnamefont {M.}~\bibnamefont
  {Kaschel}}, \bibinfo {author} {\bibfnamefont {C.}~\bibnamefont {Schweikert}},
  \bibinfo {author} {\bibfnamefont {M.}~\bibnamefont {Jetter}}, \bibinfo
  {author} {\bibfnamefont {S.~L.}\ \bibnamefont {Portalupi}}, \bibinfo {author}
  {\bibfnamefont {M.}~\bibnamefont {Berroth}},\ and\ \bibinfo {author}
  {\bibfnamefont {P.}~\bibnamefont {Michler}},\ }\bibfield  {title} {\bibinfo
  {title} {{Achieving stable fiber coupling of quantum dot telecom C-band
  single-photons to an SOI photonic device}},\ }\href
  {https://doi.org/10.1063/5.0067749} {\bibfield  {journal} {\bibinfo
  {journal} {Applied Physics Letters}\ }\textbf {\bibinfo {volume} {119}},\
  \bibinfo {pages} {211101} (\bibinfo {year} {2021})}\BibitemShut {NoStop}%
\bibitem [{\citenamefont {Sund}\ \emph {et~al.}(2023)\citenamefont {Sund},
  \citenamefont {Lomonte}, \citenamefont {Paesani}, \citenamefont {Wang},
  \citenamefont {Carolan}, \citenamefont {Bart}, \citenamefont {Wieck},
  \citenamefont {Ludwig}, \citenamefont {Midolo}, \citenamefont {Pernice},
  \citenamefont {Lodahl},\ and\ \citenamefont {Lenzini}}]{Sund.ea:2023}%
  \BibitemOpen
  \bibfield  {author} {\bibinfo {author} {\bibfnamefont {P.~I.}\ \bibnamefont
  {Sund}}, \bibinfo {author} {\bibfnamefont {E.}~\bibnamefont {Lomonte}},
  \bibinfo {author} {\bibfnamefont {S.}~\bibnamefont {Paesani}}, \bibinfo
  {author} {\bibfnamefont {Y.}~\bibnamefont {Wang}}, \bibinfo {author}
  {\bibfnamefont {J.}~\bibnamefont {Carolan}}, \bibinfo {author} {\bibfnamefont
  {N.}~\bibnamefont {Bart}}, \bibinfo {author} {\bibfnamefont {A.~D.}\
  \bibnamefont {Wieck}}, \bibinfo {author} {\bibfnamefont {A.}~\bibnamefont
  {Ludwig}}, \bibinfo {author} {\bibfnamefont {L.}~\bibnamefont {Midolo}},
  \bibinfo {author} {\bibfnamefont {W.~H.}\ \bibnamefont {Pernice}}, \bibinfo
  {author} {\bibfnamefont {P.}~\bibnamefont {Lodahl}},\ and\ \bibinfo {author}
  {\bibfnamefont {F.}~\bibnamefont {Lenzini}},\ }\bibfield  {title} {\bibinfo
  {title} {{High-speed thin-film lithium niobate quantum processor driven by a
  solid-state quantum emitter}},\ }\href
  {https://doi.org/10.1126/sciadv.adg7268} {\bibfield  {journal} {\bibinfo
  {journal} {Science Advances}\ }\textbf {\bibinfo {volume} {9}},\ \bibinfo
  {pages} {eadg7268} (\bibinfo {year} {2023})}\BibitemShut {NoStop}%
\bibitem [{\citenamefont {Arcari}\ \emph {et~al.}(2014)\citenamefont {Arcari},
  \citenamefont {Söllner}, \citenamefont {Javadi}, \citenamefont {Hansen},
  \citenamefont {Mahmoodian}, \citenamefont {Liu}, \citenamefont {Thyrrestrup},
  \citenamefont {Lee}, \citenamefont {Song}, \citenamefont {Stobbe},\ and\
  \citenamefont {Lodahl}}]{Arcari.ea:2014}%
  \BibitemOpen
  \bibfield  {author} {\bibinfo {author} {\bibfnamefont {M.}~\bibnamefont
  {Arcari}}, \bibinfo {author} {\bibfnamefont {I.}~\bibnamefont {Söllner}},
  \bibinfo {author} {\bibfnamefont {A.}~\bibnamefont {Javadi}}, \bibinfo
  {author} {\bibfnamefont {S.~L.}\ \bibnamefont {Hansen}}, \bibinfo {author}
  {\bibfnamefont {S.}~\bibnamefont {Mahmoodian}}, \bibinfo {author}
  {\bibfnamefont {J.}~\bibnamefont {Liu}}, \bibinfo {author} {\bibfnamefont
  {H.}~\bibnamefont {Thyrrestrup}}, \bibinfo {author} {\bibfnamefont
  {E.}~\bibnamefont {Lee}}, \bibinfo {author} {\bibfnamefont {J.}~\bibnamefont
  {Song}}, \bibinfo {author} {\bibfnamefont {S.}~\bibnamefont {Stobbe}},\ and\
  \bibinfo {author} {\bibfnamefont {P.}~\bibnamefont {Lodahl}},\ }\bibfield
  {title} {\bibinfo {title} {{Near-Unity Coupling Efficiency of a Quantum
  Emitter to a Photonic Crystal Waveguide}},\ }\href
  {https://doi.org/10.1103/PhysRevLett.113.093603} {\bibfield  {journal}
  {\bibinfo  {journal} {Physical Review Letters}\ }\textbf {\bibinfo {volume}
  {113}},\ \bibinfo {pages} {093603} (\bibinfo {year} {2014})}\BibitemShut
  {NoStop}%
\bibitem [{\citenamefont {Liu}\ \emph {et~al.}(2018{\natexlab{a}})\citenamefont
  {Liu}, \citenamefont {Brash}, \citenamefont {O’Hara}, \citenamefont
  {Martins}, \citenamefont {Phillips}, \citenamefont {Coles}, \citenamefont
  {Royall}, \citenamefont {Clarke}, \citenamefont {Bentham}, \citenamefont
  {Prtljaga}, \citenamefont {Itskevich}, \citenamefont {Wilson}, \citenamefont
  {Skolnick},\ and\ \citenamefont {Fox}}]{Liu.ea:2018}%
  \BibitemOpen
  \bibfield  {author} {\bibinfo {author} {\bibfnamefont {F.}~\bibnamefont
  {Liu}}, \bibinfo {author} {\bibfnamefont {A.~J.}\ \bibnamefont {Brash}},
  \bibinfo {author} {\bibfnamefont {J.}~\bibnamefont {O’Hara}}, \bibinfo
  {author} {\bibfnamefont {L.~M. P.~P.}\ \bibnamefont {Martins}}, \bibinfo
  {author} {\bibfnamefont {C.~L.}\ \bibnamefont {Phillips}}, \bibinfo {author}
  {\bibfnamefont {R.~J.}\ \bibnamefont {Coles}}, \bibinfo {author}
  {\bibfnamefont {B.}~\bibnamefont {Royall}}, \bibinfo {author} {\bibfnamefont
  {E.}~\bibnamefont {Clarke}}, \bibinfo {author} {\bibfnamefont
  {C.}~\bibnamefont {Bentham}}, \bibinfo {author} {\bibfnamefont
  {N.}~\bibnamefont {Prtljaga}}, \bibinfo {author} {\bibfnamefont {I.~E.}\
  \bibnamefont {Itskevich}}, \bibinfo {author} {\bibfnamefont {L.~R.}\
  \bibnamefont {Wilson}}, \bibinfo {author} {\bibfnamefont {M.~S.}\
  \bibnamefont {Skolnick}},\ and\ \bibinfo {author} {\bibfnamefont {A.~M.}\
  \bibnamefont {Fox}},\ }\bibfield  {title} {\bibinfo {title} {{High Purcell
  factor generation of indistinguishable on-chip single photons}},\ }\href
  {https://doi.org/10.1038/s41565-018-0188-x} {\bibfield  {journal} {\bibinfo
  {journal} {Nature Nanotechnology}\ }\textbf {\bibinfo {volume} {13}},\
  \bibinfo {pages} {835} (\bibinfo {year} {2018}{\natexlab{a}})}\BibitemShut
  {NoStop}%
\bibitem [{\citenamefont {Hepp}\ \emph {et~al.}(2018)\citenamefont {Hepp},
  \citenamefont {Bauer}, \citenamefont {Hornung}, \citenamefont {Schwartz},
  \citenamefont {Portalupi}, \citenamefont {Jetter},\ and\ \citenamefont
  {Michler}}]{Hepp.ea:2018}%
  \BibitemOpen
  \bibfield  {author} {\bibinfo {author} {\bibfnamefont {S.}~\bibnamefont
  {Hepp}}, \bibinfo {author} {\bibfnamefont {S.}~\bibnamefont {Bauer}},
  \bibinfo {author} {\bibfnamefont {F.}~\bibnamefont {Hornung}}, \bibinfo
  {author} {\bibfnamefont {M.}~\bibnamefont {Schwartz}}, \bibinfo {author}
  {\bibfnamefont {S.~L.}\ \bibnamefont {Portalupi}}, \bibinfo {author}
  {\bibfnamefont {M.}~\bibnamefont {Jetter}},\ and\ \bibinfo {author}
  {\bibfnamefont {P.}~\bibnamefont {Michler}},\ }\bibfield  {title} {\bibinfo
  {title} {{Bragg grating cavities embedded into nano-photonic waveguides for
  Purcell enhanced quantum dot emission}},\ }\href
  {https://doi.org/10.1364/OE.26.030614} {\bibfield  {journal} {\bibinfo
  {journal} {Optics Express}\ }\textbf {\bibinfo {volume} {26}},\ \bibinfo
  {pages} {30614} (\bibinfo {year} {2018})}\BibitemShut {NoStop}%
\bibitem [{\citenamefont {Łukasz Dusanowski}\ \emph
  {et~al.}(2020)\citenamefont {Łukasz Dusanowski}, \citenamefont {Köck},
  \citenamefont {Shin}, \citenamefont {Kwon}, \citenamefont {Schneider},\ and\
  \citenamefont {Höfling}}]{Dusanowski.ea:2020}%
  \BibitemOpen
  \bibfield  {author} {\bibinfo {author} {\bibnamefont {Łukasz Dusanowski}},
  \bibinfo {author} {\bibfnamefont {D.}~\bibnamefont {Köck}}, \bibinfo
  {author} {\bibfnamefont {E.}~\bibnamefont {Shin}}, \bibinfo {author}
  {\bibfnamefont {S.-H.}\ \bibnamefont {Kwon}}, \bibinfo {author}
  {\bibfnamefont {C.}~\bibnamefont {Schneider}},\ and\ \bibinfo {author}
  {\bibfnamefont {S.}~\bibnamefont {Höfling}},\ }\bibfield  {title} {\bibinfo
  {title} {{Purcell-Enhanced and Indistinguishable Single-Photon Generation
  from Quantum Dots Coupled to On-Chip Integrated Ring Resonators}},\ }\href
  {https://doi.org/10.1021/acs.nanolett.0c01771} {\bibfield  {journal}
  {\bibinfo  {journal} {Nano Letters}\ }\textbf {\bibinfo {volume} {20}},\
  \bibinfo {pages} {6357} (\bibinfo {year} {2020})}\BibitemShut {NoStop}%
\bibitem [{\citenamefont {Kim}\ \emph {et~al.}(2018)\citenamefont {Kim},
  \citenamefont {Aghaeimeibodi}, \citenamefont {Richardson}, \citenamefont
  {Leavitt},\ and\ \citenamefont {Waks}}]{Kim.ea:2018}%
  \BibitemOpen
  \bibfield  {author} {\bibinfo {author} {\bibfnamefont {J.-H.}\ \bibnamefont
  {Kim}}, \bibinfo {author} {\bibfnamefont {S.}~\bibnamefont {Aghaeimeibodi}},
  \bibinfo {author} {\bibfnamefont {C.~J.~K.}\ \bibnamefont {Richardson}},
  \bibinfo {author} {\bibfnamefont {R.~P.}\ \bibnamefont {Leavitt}},\ and\
  \bibinfo {author} {\bibfnamefont {E.}~\bibnamefont {Waks}},\ }\bibfield
  {title} {\bibinfo {title} {{Super-Radiant Emission from Quantum Dots in a
  Nanophotonic Waveguide}},\ }\href
  {https://doi.org/10.1021/acs.nanolett.8b01133} {\bibfield  {journal}
  {\bibinfo  {journal} {Nano Letters}\ }\textbf {\bibinfo {volume} {18}},\
  \bibinfo {pages} {4734} (\bibinfo {year} {2018})}\BibitemShut {NoStop}%
\bibitem [{\citenamefont {Katsumi}\ \emph {et~al.}(2020)\citenamefont
  {Katsumi}, \citenamefont {Ota}, \citenamefont {Osada}, \citenamefont
  {Tajiri}, \citenamefont {Yamaguchi}, \citenamefont {Kakuda}, \citenamefont
  {Iwamoto}, \citenamefont {Akiyama},\ and\ \citenamefont
  {Arakawa}}]{Katsumi.ea:2020}%
  \BibitemOpen
  \bibfield  {author} {\bibinfo {author} {\bibfnamefont {R.}~\bibnamefont
  {Katsumi}}, \bibinfo {author} {\bibfnamefont {Y.}~\bibnamefont {Ota}},
  \bibinfo {author} {\bibfnamefont {A.}~\bibnamefont {Osada}}, \bibinfo
  {author} {\bibfnamefont {T.}~\bibnamefont {Tajiri}}, \bibinfo {author}
  {\bibfnamefont {T.}~\bibnamefont {Yamaguchi}}, \bibinfo {author}
  {\bibfnamefont {M.}~\bibnamefont {Kakuda}}, \bibinfo {author} {\bibfnamefont
  {S.}~\bibnamefont {Iwamoto}}, \bibinfo {author} {\bibfnamefont
  {H.}~\bibnamefont {Akiyama}},\ and\ \bibinfo {author} {\bibfnamefont
  {Y.}~\bibnamefont {Arakawa}},\ }\bibfield  {title} {\bibinfo {title} {{In
  situ wavelength tuning of quantum-dot single-photon sources integrated on a
  CMOS-processed silicon waveguide}},\ }\href
  {https://doi.org/10.1063/1.5129325} {\bibfield  {journal} {\bibinfo
  {journal} {Applied Physics Letters}\ }\textbf {\bibinfo {volume} {116}},\
  \bibinfo {pages} {041103} (\bibinfo {year} {2020})}\BibitemShut {NoStop}%
\bibitem [{\citenamefont {Grim}\ \emph {et~al.}(2019)\citenamefont {Grim},
  \citenamefont {Bracker}, \citenamefont {Zalalutdinov}, \citenamefont
  {Carter}, \citenamefont {Kozen}, \citenamefont {Kim}, \citenamefont {Kim},
  \citenamefont {Mlack}, \citenamefont {Yakes}, \citenamefont {Lee},\ and\
  \citenamefont {Gammon}}]{Grim.ea:2019}%
  \BibitemOpen
  \bibfield  {author} {\bibinfo {author} {\bibfnamefont {J.~Q.}\ \bibnamefont
  {Grim}}, \bibinfo {author} {\bibfnamefont {A.~S.}\ \bibnamefont {Bracker}},
  \bibinfo {author} {\bibfnamefont {M.}~\bibnamefont {Zalalutdinov}}, \bibinfo
  {author} {\bibfnamefont {S.~G.}\ \bibnamefont {Carter}}, \bibinfo {author}
  {\bibfnamefont {A.~C.}\ \bibnamefont {Kozen}}, \bibinfo {author}
  {\bibfnamefont {M.}~\bibnamefont {Kim}}, \bibinfo {author} {\bibfnamefont
  {C.~S.}\ \bibnamefont {Kim}}, \bibinfo {author} {\bibfnamefont {J.~T.}\
  \bibnamefont {Mlack}}, \bibinfo {author} {\bibfnamefont {M.}~\bibnamefont
  {Yakes}}, \bibinfo {author} {\bibfnamefont {B.}~\bibnamefont {Lee}},\ and\
  \bibinfo {author} {\bibfnamefont {D.}~\bibnamefont {Gammon}},\ }\bibfield
  {title} {\bibinfo {title} {{Scalable in operando strain tuning in
  nanophotonic waveguides enabling three-quantum-dot superradiance}},\ }\href
  {https://doi.org/10.1038/s41563-019-0418-0} {\bibfield  {journal} {\bibinfo
  {journal} {Nature Materials}\ }\textbf {\bibinfo {volume} {18}},\ \bibinfo
  {pages} {963} (\bibinfo {year} {2019})}\BibitemShut {NoStop}%
\bibitem [{\citenamefont {Elshaari}\ \emph {et~al.}(2018)\citenamefont
  {Elshaari}, \citenamefont {Büyüközer}, \citenamefont {Zadeh},
  \citenamefont {Lettner}, \citenamefont {Zhao}, \citenamefont {Schöll},
  \citenamefont {Gyger}, \citenamefont {Reimer}, \citenamefont {Dalacu},
  \citenamefont {Poole}, \citenamefont {Jöns},\ and\ \citenamefont
  {Zwiller}}]{Elshaari.ea:2018}%
  \BibitemOpen
  \bibfield  {author} {\bibinfo {author} {\bibfnamefont {A.~W.}\ \bibnamefont
  {Elshaari}}, \bibinfo {author} {\bibfnamefont {E.}~\bibnamefont
  {Büyüközer}}, \bibinfo {author} {\bibfnamefont {I.~E.}\ \bibnamefont
  {Zadeh}}, \bibinfo {author} {\bibfnamefont {T.}~\bibnamefont {Lettner}},
  \bibinfo {author} {\bibfnamefont {P.}~\bibnamefont {Zhao}}, \bibinfo {author}
  {\bibfnamefont {E.}~\bibnamefont {Schöll}}, \bibinfo {author} {\bibfnamefont
  {S.}~\bibnamefont {Gyger}}, \bibinfo {author} {\bibfnamefont {M.~E.}\
  \bibnamefont {Reimer}}, \bibinfo {author} {\bibfnamefont {D.}~\bibnamefont
  {Dalacu}}, \bibinfo {author} {\bibfnamefont {P.~J.}\ \bibnamefont {Poole}},
  \bibinfo {author} {\bibfnamefont {K.~D.}\ \bibnamefont {Jöns}},\ and\
  \bibinfo {author} {\bibfnamefont {V.}~\bibnamefont {Zwiller}},\ }\bibfield
  {title} {\bibinfo {title} {{Strain-Tunable Quantum Integrated Photonics}},\
  }\href {https://doi.org/10.1021/acs.nanolett.8b03937} {\bibfield  {journal}
  {\bibinfo  {journal} {Nano Letters}\ }\textbf {\bibinfo {volume} {18}},\
  \bibinfo {pages} {7969} (\bibinfo {year} {2018})}\BibitemShut {NoStop}%
\bibitem [{\citenamefont {Tao}\ \emph {et~al.}(2020)\citenamefont {Tao},
  \citenamefont {Wei}, \citenamefont {Li}, \citenamefont {Ou}, \citenamefont
  {Wang}, \citenamefont {Wang}, \citenamefont {Zhang}, \citenamefont {Zhang},
  \citenamefont {Gan},\ and\ \citenamefont {Ou}}]{Tao.ea:2020}%
  \BibitemOpen
  \bibfield  {author} {\bibinfo {author} {\bibfnamefont {L.}~\bibnamefont
  {Tao}}, \bibinfo {author} {\bibfnamefont {W.}~\bibnamefont {Wei}}, \bibinfo
  {author} {\bibfnamefont {Y.}~\bibnamefont {Li}}, \bibinfo {author}
  {\bibfnamefont {W.}~\bibnamefont {Ou}}, \bibinfo {author} {\bibfnamefont
  {T.}~\bibnamefont {Wang}}, \bibinfo {author} {\bibfnamefont {C.}~\bibnamefont
  {Wang}}, \bibinfo {author} {\bibfnamefont {J.}~\bibnamefont {Zhang}},
  \bibinfo {author} {\bibfnamefont {J.}~\bibnamefont {Zhang}}, \bibinfo
  {author} {\bibfnamefont {F.}~\bibnamefont {Gan}},\ and\ \bibinfo {author}
  {\bibfnamefont {X.}~\bibnamefont {Ou}},\ }\bibfield  {title} {\bibinfo
  {title} {{On-Chip Integration of Energy-Tunable Quantum Dot Based
  Single-Photon Sources via Strain Tuning of GaAs Waveguides}},\ }\href
  {https://doi.org/10.1021/acsphotonics.0c00748} {\bibfield  {journal}
  {\bibinfo  {journal} {ACS Photonics}\ }\textbf {\bibinfo {volume} {7}},\
  \bibinfo {pages} {2723} (\bibinfo {year} {2020})}\BibitemShut {NoStop}%
\bibitem [{\citenamefont {Hepp}\ \emph {et~al.}(2020)\citenamefont {Hepp},
  \citenamefont {Hornung}, \citenamefont {Bauer}, \citenamefont {Hesselmeier},
  \citenamefont {Yuan}, \citenamefont {Jetter}, \citenamefont {Portalupi},
  \citenamefont {Rastelli},\ and\ \citenamefont {Michler}}]{Hepp.ea:2020}%
  \BibitemOpen
  \bibfield  {author} {\bibinfo {author} {\bibfnamefont {S.}~\bibnamefont
  {Hepp}}, \bibinfo {author} {\bibfnamefont {F.}~\bibnamefont {Hornung}},
  \bibinfo {author} {\bibfnamefont {S.}~\bibnamefont {Bauer}}, \bibinfo
  {author} {\bibfnamefont {E.}~\bibnamefont {Hesselmeier}}, \bibinfo {author}
  {\bibfnamefont {X.}~\bibnamefont {Yuan}}, \bibinfo {author} {\bibfnamefont
  {M.}~\bibnamefont {Jetter}}, \bibinfo {author} {\bibfnamefont {S.~L.}\
  \bibnamefont {Portalupi}}, \bibinfo {author} {\bibfnamefont {A.}~\bibnamefont
  {Rastelli}},\ and\ \bibinfo {author} {\bibfnamefont {P.}~\bibnamefont
  {Michler}},\ }\bibfield  {title} {\bibinfo {title} {{Purcell-enhanced
  single-photon emission from a strain-tunable quantum dot in a
  cavity-waveguide device}},\ }\href {https://doi.org/10.1063/5.0033213}
  {\bibfield  {journal} {\bibinfo  {journal} {Applied Physics Letters}\
  }\textbf {\bibinfo {volume} {117}},\ \bibinfo {pages} {254002} (\bibinfo
  {year} {2020})}\BibitemShut {NoStop}%
\bibitem [{\citenamefont {Petruzzella}\ \emph {et~al.}(2018)\citenamefont
  {Petruzzella}, \citenamefont {Birindelli}, \citenamefont {Pagliano},
  \citenamefont {Pellegrino}, \citenamefont {Zobenica}, \citenamefont {Li},
  \citenamefont {Linfield},\ and\ \citenamefont {Fiore}}]{Petruzzella.ea:2018}%
  \BibitemOpen
  \bibfield  {author} {\bibinfo {author} {\bibfnamefont {M.}~\bibnamefont
  {Petruzzella}}, \bibinfo {author} {\bibfnamefont {S.}~\bibnamefont
  {Birindelli}}, \bibinfo {author} {\bibfnamefont {F.~M.}\ \bibnamefont
  {Pagliano}}, \bibinfo {author} {\bibfnamefont {D.}~\bibnamefont
  {Pellegrino}}, \bibinfo {author} {\bibfnamefont {{\v{Z}}.}~\bibnamefont
  {Zobenica}}, \bibinfo {author} {\bibfnamefont {L.~H.}\ \bibnamefont {Li}},
  \bibinfo {author} {\bibfnamefont {E.~H.}\ \bibnamefont {Linfield}},\ and\
  \bibinfo {author} {\bibfnamefont {A.}~\bibnamefont {Fiore}},\ }\bibfield
  {title} {\bibinfo {title} {{Quantum photonic integrated circuits based on
  tunable dots and tunable cavities}},\ }\href
  {https://doi.org/10.1063/1.5039961} {\bibfield  {journal} {\bibinfo
  {journal} {APL Photonics}\ }\textbf {\bibinfo {volume} {3}},\ \bibinfo
  {pages} {106103} (\bibinfo {year} {2018})}\BibitemShut {NoStop}%
\bibitem [{\citenamefont {Schnauber}\ \emph {et~al.}(2021)\citenamefont
  {Schnauber}, \citenamefont {Große}, \citenamefont {Kaganskiy}, \citenamefont
  {Ott}, \citenamefont {Anikin}, \citenamefont {Schmidt}, \citenamefont
  {Rodt},\ and\ \citenamefont {Reitzenstein}}]{Schnauber.ea:2021}%
  \BibitemOpen
  \bibfield  {author} {\bibinfo {author} {\bibfnamefont {P.}~\bibnamefont
  {Schnauber}}, \bibinfo {author} {\bibfnamefont {J.}~\bibnamefont {Große}},
  \bibinfo {author} {\bibfnamefont {A.}~\bibnamefont {Kaganskiy}}, \bibinfo
  {author} {\bibfnamefont {M.}~\bibnamefont {Ott}}, \bibinfo {author}
  {\bibfnamefont {P.}~\bibnamefont {Anikin}}, \bibinfo {author} {\bibfnamefont
  {R.}~\bibnamefont {Schmidt}}, \bibinfo {author} {\bibfnamefont
  {S.}~\bibnamefont {Rodt}},\ and\ \bibinfo {author} {\bibfnamefont
  {S.}~\bibnamefont {Reitzenstein}},\ }\bibfield  {title} {\bibinfo {title}
  {{Spectral control of deterministically fabricated quantum dot waveguide
  systems using the quantum confined Stark effect}},\ }\href
  {https://doi.org/10.1063/5.0050152} {\bibfield  {journal} {\bibinfo
  {journal} {APL Photonics}\ }\textbf {\bibinfo {volume} {6}},\ \bibinfo
  {pages} {050801} (\bibinfo {year} {2021})}\BibitemShut {NoStop}%
\bibitem [{\citenamefont {Papon}\ \emph {et~al.}(2023)\citenamefont {Papon},
  \citenamefont {Wang}, \citenamefont {Uppu}, \citenamefont {Scholz},
  \citenamefont {Wieck}, \citenamefont {Ludwig}, \citenamefont {Lodahl},\ and\
  \citenamefont {Midolo}}]{Papon.ea:2023}%
  \BibitemOpen
  \bibfield  {author} {\bibinfo {author} {\bibfnamefont {C.}~\bibnamefont
  {Papon}}, \bibinfo {author} {\bibfnamefont {Y.}~\bibnamefont {Wang}},
  \bibinfo {author} {\bibfnamefont {R.}~\bibnamefont {Uppu}}, \bibinfo {author}
  {\bibfnamefont {S.}~\bibnamefont {Scholz}}, \bibinfo {author} {\bibfnamefont
  {A.}~\bibnamefont {Wieck}}, \bibinfo {author} {\bibfnamefont
  {A.}~\bibnamefont {Ludwig}}, \bibinfo {author} {\bibfnamefont
  {P.}~\bibnamefont {Lodahl}},\ and\ \bibinfo {author} {\bibfnamefont
  {L.}~\bibnamefont {Midolo}},\ }\bibfield  {title} {\bibinfo {title}
  {{Independent Operation of Two Waveguide-Integrated Quantum Emitters}},\
  }\href {https://doi.org/10.1103/PhysRevApplied.19.L061003} {\bibfield
  {journal} {\bibinfo  {journal} {Physical Review Applied}\ }\textbf {\bibinfo
  {volume} {19}},\ \bibinfo {pages} {L061003} (\bibinfo {year}
  {2023})}\BibitemShut {NoStop}%
\bibitem [{\citenamefont {Łukasz Dusanowski}\ \emph
  {et~al.}(2023)\citenamefont {Łukasz Dusanowski}, \citenamefont {Köck},
  \citenamefont {Schneider},\ and\ \citenamefont
  {Höfling}}]{Dusanowski.ea:2023}%
  \BibitemOpen
  \bibfield  {author} {\bibinfo {author} {\bibnamefont {Łukasz Dusanowski}},
  \bibinfo {author} {\bibfnamefont {D.}~\bibnamefont {Köck}}, \bibinfo
  {author} {\bibfnamefont {C.}~\bibnamefont {Schneider}},\ and\ \bibinfo
  {author} {\bibfnamefont {S.}~\bibnamefont {Höfling}},\ }\bibfield  {title}
  {\bibinfo {title} {{On-Chip Hong–Ou–Mandel Interference from Separate
  Quantum Dot Emitters in an Integrated Circuit}},\ }\href
  {https://doi.org/10.1021/acsphotonics.3c00679} {\bibfield  {journal}
  {\bibinfo  {journal} {ACS Photonics}\ }\textbf {\bibinfo {volume} {10}},\
  \bibinfo {pages} {2941} (\bibinfo {year} {2023})}\BibitemShut {NoStop}%
\bibitem [{\citenamefont {Reindl}\ \emph {et~al.}(2019)\citenamefont {Reindl},
  \citenamefont {Weber}, \citenamefont {Huber}, \citenamefont {Schimpf},
  \citenamefont {da~Silva}, \citenamefont {Portalupi}, \citenamefont {Trotta},
  \citenamefont {Michler},\ and\ \citenamefont {Rastelli}}]{Reindl.ea:2019}%
  \BibitemOpen
  \bibfield  {author} {\bibinfo {author} {\bibfnamefont {M.}~\bibnamefont
  {Reindl}}, \bibinfo {author} {\bibfnamefont {J.~H.}\ \bibnamefont {Weber}},
  \bibinfo {author} {\bibfnamefont {D.}~\bibnamefont {Huber}}, \bibinfo
  {author} {\bibfnamefont {C.}~\bibnamefont {Schimpf}}, \bibinfo {author}
  {\bibfnamefont {S.~F.~C.}\ \bibnamefont {da~Silva}}, \bibinfo {author}
  {\bibfnamefont {S.~L.}\ \bibnamefont {Portalupi}}, \bibinfo {author}
  {\bibfnamefont {R.}~\bibnamefont {Trotta}}, \bibinfo {author} {\bibfnamefont
  {P.}~\bibnamefont {Michler}},\ and\ \bibinfo {author} {\bibfnamefont
  {A.}~\bibnamefont {Rastelli}},\ }\bibfield  {title} {\bibinfo {title}
  {{Highly indistinguishable single photons from incoherently excited quantum
  dots}},\ }\href {https://doi.org/10.1103/PhysRevB.100.155420} {\bibfield
  {journal} {\bibinfo  {journal} {Physical Review B}\ }\textbf {\bibinfo
  {volume} {100}},\ \bibinfo {pages} {155420} (\bibinfo {year}
  {2019})}\BibitemShut {NoStop}%
\bibitem [{\citenamefont {Schöll}\ \emph {et~al.}(2019)\citenamefont
  {Schöll}, \citenamefont {Hanschke}, \citenamefont {Schweickert},
  \citenamefont {Zeuner}, \citenamefont {Reindl}, \citenamefont {da~Silva},
  \citenamefont {Lettner}, \citenamefont {Trotta}, \citenamefont {Finley},
  \citenamefont {Müller}, \citenamefont {Rastelli}, \citenamefont {Zwiller},\
  and\ \citenamefont {Jöns}}]{Schoell.ea:2019}%
  \BibitemOpen
  \bibfield  {author} {\bibinfo {author} {\bibfnamefont {E.}~\bibnamefont
  {Schöll}}, \bibinfo {author} {\bibfnamefont {L.}~\bibnamefont {Hanschke}},
  \bibinfo {author} {\bibfnamefont {L.}~\bibnamefont {Schweickert}}, \bibinfo
  {author} {\bibfnamefont {K.~D.}\ \bibnamefont {Zeuner}}, \bibinfo {author}
  {\bibfnamefont {M.}~\bibnamefont {Reindl}}, \bibinfo {author} {\bibfnamefont
  {S.~F.~C.}\ \bibnamefont {da~Silva}}, \bibinfo {author} {\bibfnamefont
  {T.}~\bibnamefont {Lettner}}, \bibinfo {author} {\bibfnamefont
  {R.}~\bibnamefont {Trotta}}, \bibinfo {author} {\bibfnamefont {J.~J.}\
  \bibnamefont {Finley}}, \bibinfo {author} {\bibfnamefont {K.}~\bibnamefont
  {Müller}}, \bibinfo {author} {\bibfnamefont {A.}~\bibnamefont {Rastelli}},
  \bibinfo {author} {\bibfnamefont {V.}~\bibnamefont {Zwiller}},\ and\ \bibinfo
  {author} {\bibfnamefont {K.~D.}\ \bibnamefont {Jöns}},\ }\bibfield  {title}
  {\bibinfo {title} {{Resonance Fluorescence of GaAs Quantum Dots with
  Near-Unity Photon Indistinguishability}},\ }\href
  {https://doi.org/10.1021/acs.nanolett.8b05132} {\bibfield  {journal}
  {\bibinfo  {journal} {Nano Letters}\ }\textbf {\bibinfo {volume} {19}},\
  \bibinfo {pages} {2404} (\bibinfo {year} {2019})}\BibitemShut {NoStop}%
\bibitem [{\citenamefont {da~Silva}\ \emph {et~al.}(2021)\citenamefont
  {da~Silva}, \citenamefont {Undeutsch}, \citenamefont {Lehner}, \citenamefont
  {Manna}, \citenamefont {Krieger}, \citenamefont {Reindl}, \citenamefont
  {Schimpf}, \citenamefont {Trotta},\ and\ \citenamefont
  {Rastelli}}]{daSilva.ea:2021}%
  \BibitemOpen
  \bibfield  {author} {\bibinfo {author} {\bibfnamefont {S.~F.~C.}\
  \bibnamefont {da~Silva}}, \bibinfo {author} {\bibfnamefont {G.}~\bibnamefont
  {Undeutsch}}, \bibinfo {author} {\bibfnamefont {B.}~\bibnamefont {Lehner}},
  \bibinfo {author} {\bibfnamefont {S.}~\bibnamefont {Manna}}, \bibinfo
  {author} {\bibfnamefont {T.~M.}\ \bibnamefont {Krieger}}, \bibinfo {author}
  {\bibfnamefont {M.}~\bibnamefont {Reindl}}, \bibinfo {author} {\bibfnamefont
  {C.}~\bibnamefont {Schimpf}}, \bibinfo {author} {\bibfnamefont
  {R.}~\bibnamefont {Trotta}},\ and\ \bibinfo {author} {\bibfnamefont
  {A.}~\bibnamefont {Rastelli}},\ }\bibfield  {title} {\bibinfo {title} {{GaAs
  quantum dots grown by droplet etching epitaxy as quantum light sources}},\
  }\href {https://doi.org/10.1063/5.0057070} {\bibfield  {journal} {\bibinfo
  {journal} {Applied Physics Letters}\ }\textbf {\bibinfo {volume} {119}},\
  \bibinfo {pages} {120502} (\bibinfo {year} {2021})}\BibitemShut {NoStop}%
\bibitem [{\citenamefont {Huber}\ \emph {et~al.}(2018)\citenamefont {Huber},
  \citenamefont {Reindl}, \citenamefont {da~Silva}, \citenamefont {Schimpf},
  \citenamefont {Martín-Sánchez}, \citenamefont {Huang}, \citenamefont
  {Piredda}, \citenamefont {Edlinger}, \citenamefont {Rastelli},\ and\
  \citenamefont {Trotta}}]{Huber.ea:2018}%
  \BibitemOpen
  \bibfield  {author} {\bibinfo {author} {\bibfnamefont {D.}~\bibnamefont
  {Huber}}, \bibinfo {author} {\bibfnamefont {M.}~\bibnamefont {Reindl}},
  \bibinfo {author} {\bibfnamefont {S.~F.~C.}\ \bibnamefont {da~Silva}},
  \bibinfo {author} {\bibfnamefont {C.}~\bibnamefont {Schimpf}}, \bibinfo
  {author} {\bibfnamefont {J.}~\bibnamefont {Martín-Sánchez}}, \bibinfo
  {author} {\bibfnamefont {H.}~\bibnamefont {Huang}}, \bibinfo {author}
  {\bibfnamefont {G.}~\bibnamefont {Piredda}}, \bibinfo {author} {\bibfnamefont
  {J.}~\bibnamefont {Edlinger}}, \bibinfo {author} {\bibfnamefont
  {A.}~\bibnamefont {Rastelli}},\ and\ \bibinfo {author} {\bibfnamefont
  {R.}~\bibnamefont {Trotta}},\ }\bibfield  {title} {\bibinfo {title}
  {{Strain-Tunable GaAs Quantum Dot: A Nearly Dephasing-Free Source of
  Entangled Photon Pairs on Demand}},\ }\href
  {https://doi.org/10.1103/PhysRevLett.121.033902} {\bibfield  {journal}
  {\bibinfo  {journal} {Physical Review Letters}\ }\textbf {\bibinfo {volume}
  {121}},\ \bibinfo {pages} {033902} (\bibinfo {year} {2018})}\BibitemShut
  {NoStop}%
\bibitem [{\citenamefont {Huang}\ \emph {et~al.}(2021)\citenamefont {Huang},
  \citenamefont {Csontosová}, \citenamefont {Manna}, \citenamefont {Huo},
  \citenamefont {Trotta}, \citenamefont {Rastelli},\ and\ \citenamefont
  {Klenovský}}]{Huang.ea:2021}%
  \BibitemOpen
  \bibfield  {author} {\bibinfo {author} {\bibfnamefont {H.}~\bibnamefont
  {Huang}}, \bibinfo {author} {\bibfnamefont {D.}~\bibnamefont {Csontosová}},
  \bibinfo {author} {\bibfnamefont {S.}~\bibnamefont {Manna}}, \bibinfo
  {author} {\bibfnamefont {Y.}~\bibnamefont {Huo}}, \bibinfo {author}
  {\bibfnamefont {R.}~\bibnamefont {Trotta}}, \bibinfo {author} {\bibfnamefont
  {A.}~\bibnamefont {Rastelli}},\ and\ \bibinfo {author} {\bibfnamefont
  {P.}~\bibnamefont {Klenovský}},\ }\bibfield  {title} {\bibinfo {title}
  {{Electric field induced tuning of electronic correlation in weakly confining
  quantum dots}},\ }\href {https://doi.org/10.1103/PhysRevB.104.165401}
  {\bibfield  {journal} {\bibinfo  {journal} {Physical Review B}\ }\textbf
  {\bibinfo {volume} {104}},\ \bibinfo {pages} {165401} (\bibinfo {year}
  {2021})}\BibitemShut {NoStop}%
\bibitem [{\citenamefont {Zhai}\ \emph {et~al.}(2020)\citenamefont {Zhai},
  \citenamefont {Löbl}, \citenamefont {Nguyen}, \citenamefont {Ritzmann},
  \citenamefont {Javadi}, \citenamefont {Spinnler}, \citenamefont {Wieck},
  \citenamefont {Ludwig},\ and\ \citenamefont {Warburton}}]{Zhai.ea:2020}%
  \BibitemOpen
  \bibfield  {author} {\bibinfo {author} {\bibfnamefont {L.}~\bibnamefont
  {Zhai}}, \bibinfo {author} {\bibfnamefont {M.~C.}\ \bibnamefont {Löbl}},
  \bibinfo {author} {\bibfnamefont {G.~N.}\ \bibnamefont {Nguyen}}, \bibinfo
  {author} {\bibfnamefont {J.}~\bibnamefont {Ritzmann}}, \bibinfo {author}
  {\bibfnamefont {A.}~\bibnamefont {Javadi}}, \bibinfo {author} {\bibfnamefont
  {C.}~\bibnamefont {Spinnler}}, \bibinfo {author} {\bibfnamefont {A.~D.}\
  \bibnamefont {Wieck}}, \bibinfo {author} {\bibfnamefont {A.}~\bibnamefont
  {Ludwig}},\ and\ \bibinfo {author} {\bibfnamefont {R.~J.}\ \bibnamefont
  {Warburton}},\ }\bibfield  {title} {\bibinfo {title} {{Low-noise GaAs quantum
  dots for quantum photonics}},\ }\href
  {https://doi.org/10.1038/s41467-020-18625-z} {\bibfield  {journal} {\bibinfo
  {journal} {Nature Communications}\ }\textbf {\bibinfo {volume} {11}},\
  \bibinfo {pages} {4745} (\bibinfo {year} {2020})}\BibitemShut {NoStop}%
\bibitem [{\citenamefont {Zhai}\ \emph {et~al.}(2022)\citenamefont {Zhai},
  \citenamefont {Nguyen}, \citenamefont {Spinnler}, \citenamefont {Ritzmann},
  \citenamefont {Löbl}, \citenamefont {Wieck}, \citenamefont {Ludwig},
  \citenamefont {Javadi},\ and\ \citenamefont {Warburton}}]{Zhai.ea:2022}%
  \BibitemOpen
  \bibfield  {author} {\bibinfo {author} {\bibfnamefont {L.}~\bibnamefont
  {Zhai}}, \bibinfo {author} {\bibfnamefont {G.~N.}\ \bibnamefont {Nguyen}},
  \bibinfo {author} {\bibfnamefont {C.}~\bibnamefont {Spinnler}}, \bibinfo
  {author} {\bibfnamefont {J.}~\bibnamefont {Ritzmann}}, \bibinfo {author}
  {\bibfnamefont {M.~C.}\ \bibnamefont {Löbl}}, \bibinfo {author}
  {\bibfnamefont {A.~D.}\ \bibnamefont {Wieck}}, \bibinfo {author}
  {\bibfnamefont {A.}~\bibnamefont {Ludwig}}, \bibinfo {author} {\bibfnamefont
  {A.}~\bibnamefont {Javadi}},\ and\ \bibinfo {author} {\bibfnamefont {R.~J.}\
  \bibnamefont {Warburton}},\ }\bibfield  {title} {\bibinfo {title} {{Quantum
  interference of identical photons from remote GaAs quantum dots}},\ }\href
  {https://doi.org/10.1038/s41565-022-01131-2} {\bibfield  {journal} {\bibinfo
  {journal} {Nature Nanotechnology}\ }\textbf {\bibinfo {volume} {17}},\
  \bibinfo {pages} {829} (\bibinfo {year} {2022})}\BibitemShut {NoStop}%
\bibitem [{\citenamefont {Schwartz}\ \emph {et~al.}(2016)\citenamefont
  {Schwartz}, \citenamefont {Rengstl}, \citenamefont {Herzog}, \citenamefont
  {Paul}, \citenamefont {Kettler}, \citenamefont {Portalupi}, \citenamefont
  {Jetter},\ and\ \citenamefont {Michler}}]{Schwartz.ea:2016}%
  \BibitemOpen
  \bibfield  {author} {\bibinfo {author} {\bibfnamefont {M.}~\bibnamefont
  {Schwartz}}, \bibinfo {author} {\bibfnamefont {U.}~\bibnamefont {Rengstl}},
  \bibinfo {author} {\bibfnamefont {T.}~\bibnamefont {Herzog}}, \bibinfo
  {author} {\bibfnamefont {M.}~\bibnamefont {Paul}}, \bibinfo {author}
  {\bibfnamefont {J.}~\bibnamefont {Kettler}}, \bibinfo {author} {\bibfnamefont
  {S.~L.}\ \bibnamefont {Portalupi}}, \bibinfo {author} {\bibfnamefont
  {M.}~\bibnamefont {Jetter}},\ and\ \bibinfo {author} {\bibfnamefont
  {P.}~\bibnamefont {Michler}},\ }\bibfield  {title} {\bibinfo {title}
  {{Generation, guiding and splitting of triggered single photons from a
  resonantly excited quantum dot in a photonic circuit}},\ }\href
  {https://doi.org/10.1364/OE.24.003089} {\bibfield  {journal} {\bibinfo
  {journal} {Optics Express}\ }\textbf {\bibinfo {volume} {24}},\ \bibinfo
  {pages} {3089} (\bibinfo {year} {2016})}\BibitemShut {NoStop}%
\bibitem [{\citenamefont {Makhonin}\ \emph {et~al.}(2014)\citenamefont
  {Makhonin}, \citenamefont {Dixon}, \citenamefont {Coles}, \citenamefont
  {Royall}, \citenamefont {Luxmoore}, \citenamefont {Clarke}, \citenamefont
  {Hugues}, \citenamefont {Skolnick},\ and\ \citenamefont
  {Fox}}]{Makhonin.ea:2014}%
  \BibitemOpen
  \bibfield  {author} {\bibinfo {author} {\bibfnamefont {M.~N.}\ \bibnamefont
  {Makhonin}}, \bibinfo {author} {\bibfnamefont {J.~E.}\ \bibnamefont {Dixon}},
  \bibinfo {author} {\bibfnamefont {R.~J.}\ \bibnamefont {Coles}}, \bibinfo
  {author} {\bibfnamefont {B.}~\bibnamefont {Royall}}, \bibinfo {author}
  {\bibfnamefont {I.~J.}\ \bibnamefont {Luxmoore}}, \bibinfo {author}
  {\bibfnamefont {E.}~\bibnamefont {Clarke}}, \bibinfo {author} {\bibfnamefont
  {M.}~\bibnamefont {Hugues}}, \bibinfo {author} {\bibfnamefont {M.~S.}\
  \bibnamefont {Skolnick}},\ and\ \bibinfo {author} {\bibfnamefont {A.~M.}\
  \bibnamefont {Fox}},\ }\bibfield  {title} {\bibinfo {title} {{Waveguide
  Coupled Resonance Fluorescence from On-Chip Quantum Emitter}},\ }\href
  {https://doi.org/10.1021/nl5032937} {\bibfield  {journal} {\bibinfo
  {journal} {Nano Letters}\ }\textbf {\bibinfo {volume} {14}},\ \bibinfo
  {pages} {6997} (\bibinfo {year} {2014})}\BibitemShut {NoStop}%
\bibitem [{\citenamefont {Huber}\ \emph {et~al.}(2020)\citenamefont {Huber},
  \citenamefont {Davanco}, \citenamefont {Müller}, \citenamefont {Shuai},
  \citenamefont {Gazzano},\ and\ \citenamefont {Solomon}}]{Huber.ea:2020}%
  \BibitemOpen
  \bibfield  {author} {\bibinfo {author} {\bibfnamefont {T.}~\bibnamefont
  {Huber}}, \bibinfo {author} {\bibfnamefont {M.}~\bibnamefont {Davanco}},
  \bibinfo {author} {\bibfnamefont {M.}~\bibnamefont {Müller}}, \bibinfo
  {author} {\bibfnamefont {Y.}~\bibnamefont {Shuai}}, \bibinfo {author}
  {\bibfnamefont {O.}~\bibnamefont {Gazzano}},\ and\ \bibinfo {author}
  {\bibfnamefont {G.~S.}\ \bibnamefont {Solomon}},\ }\bibfield  {title}
  {\bibinfo {title} {{Filter-free single-photon quantum dot resonance
  fluorescence in an integrated cavity-waveguide device}},\ }\href
  {https://doi.org/10.1364/OPTICA.382273} {\bibfield  {journal} {\bibinfo
  {journal} {Optica}\ }\textbf {\bibinfo {volume} {7}},\ \bibinfo {pages} {380}
  (\bibinfo {year} {2020})}\BibitemShut {NoStop}%
\bibitem [{\citenamefont {Liu}\ \emph {et~al.}(2018{\natexlab{b}})\citenamefont
  {Liu}, \citenamefont {Konthasinghe}, \citenamefont {Davanço}, \citenamefont
  {Lawall}, \citenamefont {Anant}, \citenamefont {Verma}, \citenamefont
  {Mirin}, \citenamefont {Nam}, \citenamefont {Song}, \citenamefont {Ma},
  \citenamefont {Chen}, \citenamefont {Ni}, \citenamefont {Niu},\ and\
  \citenamefont {Srinivasan}}]{Liu.ea:2018_2}%
  \BibitemOpen
  \bibfield  {author} {\bibinfo {author} {\bibfnamefont {J.}~\bibnamefont
  {Liu}}, \bibinfo {author} {\bibfnamefont {K.}~\bibnamefont {Konthasinghe}},
  \bibinfo {author} {\bibfnamefont {M.}~\bibnamefont {Davanço}}, \bibinfo
  {author} {\bibfnamefont {J.}~\bibnamefont {Lawall}}, \bibinfo {author}
  {\bibfnamefont {V.}~\bibnamefont {Anant}}, \bibinfo {author} {\bibfnamefont
  {V.}~\bibnamefont {Verma}}, \bibinfo {author} {\bibfnamefont
  {R.}~\bibnamefont {Mirin}}, \bibinfo {author} {\bibfnamefont {S.~W.}\
  \bibnamefont {Nam}}, \bibinfo {author} {\bibfnamefont {J.~D.}\ \bibnamefont
  {Song}}, \bibinfo {author} {\bibfnamefont {B.}~\bibnamefont {Ma}}, \bibinfo
  {author} {\bibfnamefont {Z.~S.}\ \bibnamefont {Chen}}, \bibinfo {author}
  {\bibfnamefont {H.~Q.}\ \bibnamefont {Ni}}, \bibinfo {author} {\bibfnamefont
  {Z.~C.}\ \bibnamefont {Niu}},\ and\ \bibinfo {author} {\bibfnamefont
  {K.}~\bibnamefont {Srinivasan}},\ }\bibfield  {title} {\bibinfo {title}
  {{Single Self-Assembled InAs/GaAs Quantum Dots in Photonic Nanostructures:
  The Role of Nanofabrication}},\ }\href
  {https://doi.org/10.1103/PhysRevApplied.9.064019} {\bibfield  {journal}
  {\bibinfo  {journal} {Physical Review Applied}\ }\textbf {\bibinfo {volume}
  {9}},\ \bibinfo {pages} {064019} (\bibinfo {year}
  {2018}{\natexlab{b}})}\BibitemShut {NoStop}%
\bibitem [{\citenamefont {Manna}\ \emph {et~al.}(2020)\citenamefont {Manna},
  \citenamefont {Huang}, \citenamefont {da~Silva}, \citenamefont {Schimpf},
  \citenamefont {Rota}, \citenamefont {Lehner}, \citenamefont {Reindl},
  \citenamefont {Trotta},\ and\ \citenamefont {Rastelli}}]{Manna.ea:2020}%
  \BibitemOpen
  \bibfield  {author} {\bibinfo {author} {\bibfnamefont {S.}~\bibnamefont
  {Manna}}, \bibinfo {author} {\bibfnamefont {H.}~\bibnamefont {Huang}},
  \bibinfo {author} {\bibfnamefont {S.~F.~C.}\ \bibnamefont {da~Silva}},
  \bibinfo {author} {\bibfnamefont {C.}~\bibnamefont {Schimpf}}, \bibinfo
  {author} {\bibfnamefont {M.~B.}\ \bibnamefont {Rota}}, \bibinfo {author}
  {\bibfnamefont {B.}~\bibnamefont {Lehner}}, \bibinfo {author} {\bibfnamefont
  {M.}~\bibnamefont {Reindl}}, \bibinfo {author} {\bibfnamefont
  {R.}~\bibnamefont {Trotta}},\ and\ \bibinfo {author} {\bibfnamefont
  {A.}~\bibnamefont {Rastelli}},\ }\bibfield  {title} {\bibinfo {title}
  {{Surface passivation and oxide encapsulation to improve optical properties
  of a single GaAs quantum dot close to the surface}},\ }\href
  {https://doi.org/10.1016/j.apsusc.2020.147360} {\bibfield  {journal}
  {\bibinfo  {journal} {Applied Surface Science}\ }\textbf {\bibinfo {volume}
  {532}},\ \bibinfo {pages} {147360} (\bibinfo {year} {2020})}\BibitemShut
  {NoStop}%
\bibitem [{\citenamefont {Pregnolato}\ \emph {et~al.}(2020)\citenamefont
  {Pregnolato}, \citenamefont {Chu}, \citenamefont {Schröder}, \citenamefont
  {Schott}, \citenamefont {Wieck}, \citenamefont {Ludwig}, \citenamefont
  {Lodahl},\ and\ \citenamefont {Rotenberg}}]{Pregnolato.ea:2020}%
  \BibitemOpen
  \bibfield  {author} {\bibinfo {author} {\bibfnamefont {T.}~\bibnamefont
  {Pregnolato}}, \bibinfo {author} {\bibfnamefont {X.-L.}\ \bibnamefont {Chu}},
  \bibinfo {author} {\bibfnamefont {T.}~\bibnamefont {Schröder}}, \bibinfo
  {author} {\bibfnamefont {R.}~\bibnamefont {Schott}}, \bibinfo {author}
  {\bibfnamefont {A.~D.}\ \bibnamefont {Wieck}}, \bibinfo {author}
  {\bibfnamefont {A.}~\bibnamefont {Ludwig}}, \bibinfo {author} {\bibfnamefont
  {P.}~\bibnamefont {Lodahl}},\ and\ \bibinfo {author} {\bibfnamefont
  {N.}~\bibnamefont {Rotenberg}},\ }\bibfield  {title} {\bibinfo {title}
  {{Deterministic positioning of nanophotonic waveguides around single
  self-assembled quantum dots}},\ }\href {https://doi.org/10.1063/1.5117888}
  {\bibfield  {journal} {\bibinfo  {journal} {APL Photonics}\ }\textbf
  {\bibinfo {volume} {5}},\ \bibinfo {pages} {086101} (\bibinfo {year}
  {2020})}\BibitemShut {NoStop}%
\bibitem [{\citenamefont {Li}\ \emph {et~al.}(2023)\citenamefont {Li},
  \citenamefont {Yang}, \citenamefont {Schall}, \citenamefont {von Helversen},
  \citenamefont {Palekar}, \citenamefont {Liu}, \citenamefont {Roche},
  \citenamefont {Rodt}, \citenamefont {Ni}, \citenamefont {Zhang},
  \citenamefont {Niu},\ and\ \citenamefont {Reitzenstein}}]{Li.ea:2023}%
  \BibitemOpen
  \bibfield  {author} {\bibinfo {author} {\bibfnamefont {S.}~\bibnamefont
  {Li}}, \bibinfo {author} {\bibfnamefont {Y.}~\bibnamefont {Yang}}, \bibinfo
  {author} {\bibfnamefont {J.}~\bibnamefont {Schall}}, \bibinfo {author}
  {\bibfnamefont {M.}~\bibnamefont {von Helversen}}, \bibinfo {author}
  {\bibfnamefont {C.}~\bibnamefont {Palekar}}, \bibinfo {author} {\bibfnamefont
  {H.}~\bibnamefont {Liu}}, \bibinfo {author} {\bibfnamefont {L.}~\bibnamefont
  {Roche}}, \bibinfo {author} {\bibfnamefont {S.}~\bibnamefont {Rodt}},
  \bibinfo {author} {\bibfnamefont {H.}~\bibnamefont {Ni}}, \bibinfo {author}
  {\bibfnamefont {Y.}~\bibnamefont {Zhang}}, \bibinfo {author} {\bibfnamefont
  {Z.}~\bibnamefont {Niu}},\ and\ \bibinfo {author} {\bibfnamefont
  {S.}~\bibnamefont {Reitzenstein}},\ }\bibfield  {title} {\bibinfo {title}
  {{Scalable Deterministic Integration of Two Quantum Dots into an On-Chip
  Quantum Circuit}},\ }\href {https://doi.org/10.1021/acsphotonics.3c00547}
  {\bibfield  {journal} {\bibinfo  {journal} {ACS Photonics}\ }\textbf
  {\bibinfo {volume} {10}},\ \bibinfo {pages} {2846} (\bibinfo {year}
  {2023})}\BibitemShut {NoStop}%
\end{thebibliography}
%
\end{document}